\documentclass[aps,prb,twocolumn,longbibliography,superscriptaddress,amsfonts,citeautoscript,floatfix,maxnames=4,10pt]{revtex4-2}
\usepackage[utf8]{inputenc}
\usepackage[english]{babel}
\usepackage{graphicx}
\usepackage{float}
\usepackage{bm}
\usepackage{xcolor}
\usepackage{epstopdf}
\usepackage{amsmath} 
\usepackage{amssymb}
\usepackage{epstopdf}
\usepackage[normalem]{ulem} %strikethrough \sout 
\usepackage{enumerate}
\usepackage{cases}

\usepackage[urlcolor=blue,colorlinks=true,citecolor=red,linkcolor=red,pdfstartview={FitH},bookmarks=false]{hyperref}

\newcommand{\BZ}{{V_{\rm Z}}}
\newcommand{\Bcpm}{{V_{\pm}}}
\newcommand{\BcpmAlpha}{{V_{\pm,\alpha}}}
\newcommand{\Bcp}{{V_{+}}}
\newcommand{\Bcm}{{V_{-}}}
\newcommand{\DeltaT}{{\Delta_{\rm T}}}
\newcommand{\DeltaTalpha}{{\Delta_{\rm T}^{\alpha}}}
\newcommand{\DeltaSC}{{\Delta_{\rm SC}}}
\newcommand{\DeltaQC}{{\Delta_{\rm QC}}}

\graphicspath{{figs/}{./figs/}{.}}
\usepackage{lipsum}

\DeclareMathOperator{\sgn}{sgn}

\begin{document}

\title{Topological superconductivity in Fibonacci quasicrystals}

\author{Aksel Kobiałka}
\email[e-mail:]{aksel.kobialka@physics.uu.se}
\affiliation{Department of Physics and Astronomy, Uppsala University, Box 516, S-751 20 Uppsala, Sweden}

\author{Oladunjoye A. Awoga}
\email[e-mail:]{oladunjoye.awoga@ftf.lth.se}
\affiliation{Solid State Physics and NanoLund, Lund University, Box 118, 22100 Lund, Sweden}

\author{Martin Leijnse}
\affiliation{Solid State Physics and NanoLund, Lund University, Box 118, 22100 Lund, Sweden}

\author{Tadeusz Doma\'nski}
%\email[e-mail:]{doman@kft.umcs.lublin.pl}
\affiliation{Institute of Physics, M. Curie-Sk\l{}odowska University, 20-031 Lublin, Poland}

\author{Patric Holmvall}
\email[e-mail:]{patric.holmvall@physics.uu.se}
\affiliation{Department of Physics and Astronomy, Uppsala University, Box 516, S-751 20 Uppsala, Sweden}

\author{Annica M. Black-Schaffer}
%\email[e-mail:]{annica.black-schaffer@physics.uu.se}
\affiliation{Department of Physics and Astronomy, Uppsala University, Box 516, S-751 20 Uppsala, Sweden}
\date{\today}   

%%%%%%%%%%%%%%%%%%%%%%%%%%%%%%%%%%%%%%%%%%%%%%%%%%
\begin{abstract}
We investigate the properties of a Fibonacci quasicrystal (QC) arrangement of a one-dimensional topological superconductor, such as a magnetic atom chain deposited on a superconducting surface.
We uncover a general mutually exclusive competition between the QC properties and the topological superconducting phase with Majorana bound states (MBS):  there are no MBS inside the QC gaps and the MBS never behaves as QC subgap states, and likewise, no critical, or winding, QC subgap states exist inside the topological superconducting gaps.
Surprisingly, despite this competition, we find that the QC is still highly beneficial for realizing topological superconductivity with MBS. It both leads to additional large nontrivial regions with MBS in parameter space, that are topologically trivial in  crystalline systems, and increases the topological gap protecting the MBS. We also find that shorter approximants of the Fibonacci QC display the largest benefits.
As a consequence, our results promote QCs, and especially their short approximants, as an appealing platform for improved experimental possibilities to realize MBS as well as generally highlights the fundamental interplay between different topologies.
\end{abstract}

\maketitle

%%%%%%%%%%%%%%%%%%%%%%%%%%%%%%%%%%%%%%%%%%%%%%%%%%
\section{Introduction}
\label{sec.int}
%%%%%%%%%%%%%%%%%%%%%%%%%%%%%%%%%%%%%%%%%%%%%%%%%%

Topological superconductivity is a captivating phenomenon, including allowing for the existence of topologically protected zero-energy subgap states at the system edges.
These states become Majorana bound states (MBS) in effective spinless $p$-wave superconductors, for example generated by a combination of spin-orbit coupling (SOC), magnetism, and conventional $s$-wave superconductivity
~\cite{kitaev.01,ivanov.01,oreg.refael.10,lutchyn.sau.10,Leijnse2012Introduction,menard.mesaros.19}.
Beyond an intrinsic fascination with MBS, the current high interest in topological superconductors also stems from their possible application in quantum computation, as MBS are protected from external disturbances by their nonlocality as long as the topological gap is not closed, and can thus, in principle, retain quantum information indefinitely~\cite{ryu.hatsugai.02,alicea.12,aasen.hell.16,sedlmayr.kaladzhyan.17,flensberg.oppen.21,scheppe.pak.22}.

MBS have already been theoretically studied in a wide class of configurations, including in semiconducting-superconducting hybrid nanostructures~\cite{stanescu.lutchyn.11,ptok.kobialka.17,lutchyn.bakkers.18}, quantum Hall systems~\cite{qi.wu.06,wang.zhou.15}, Josephson junctions~\cite{hell.17,cayao.17,pientka.17,lesser.22,baldo.23,xie.23}, chains of magnetic atoms or Yu-Shiba-Rusinov states~\cite{ balatsky.vekhter.06,ji.zhang.08, nadjperge.drozdov.13,
brydon.sarma.15,ptok.glodzik.17,choi.rubio.17,Awoga2017Disorder,choi.lorente.19,theiler.bjornson.19,teixeira.kuzmanovski.20, awoga2023controlling}, and as corner states in higher-dimensional topological superconductors~\cite{wang.lin.18,pahomi.sigrist.20} or in systems with antiferromagnetic~\cite{ezawa.15,kobialka.sedlmayr.20} or skyrmion order~\cite{poyhonen.weststrom.16,yang.stano.16,garnier.mesaros.19,mascot.bedow.21,diaz.klinovaja.21,maeland.sudbo.23}.
Possible signatures of MBS have been experimentally identified in various setups~\cite{rokhinson.liu.12,wiedenmann.bocquillon.16,bocquillon.deacon.17,woerkom.proutski.17,ren.pientka.19,dartiailh.mayer.21,bai.wei.22,aghaee.23,mourik.zuo.12,finck.vanharlingen.13,deng.vaitiekenas.16,deng.vaitiekenas.18,vaitiekenas.liu.21,nichele.drachmann.17,fornieri.whiticar.19,ren.19,banerjee.lesser.23,nadjperge.drozdov.14,pawlak.kisiel.16,choi.robles.16,kim.palaciomorales.18,wang.wiebe.21,schneider2021topological,schneider.beck.22,crawford.mascot.22, beck.schneider.23,beck.njari.23,crawford.mascot.23}, but it has also become clear that trivial electronic states, especially Andreev bound states appearing accidentally around zero energy, can give rise to  similar effects~\cite{huang.setiawan.18,ji.wen.18,reeg.dmytruk.18,Awoga2019Supercurrent,kayyalha.xiao.20,frolov.23,hess.legg.23,Awoga2023Mitigating}.

The search for suitable materials and setups for the creation and manipulation of MBS is still ongoing, with their emergent nature allowing for a combination of various approaches in order to reach the goal of stability.
Taking advantage of recent developments in experimental fabrication of chains of magnetic atoms deposited on a superconducting surface, clearly allowing for atomic manipulations~\cite{nadjperge.drozdov.14,pawlak.kisiel.16,choi.robles.16,choi.rubio.17,kim.palaciomorales.18,choi.lorente.19,schneider2021topological,schneider.beck.22, crawford.mascot.22, beck.schneider.23,beck.njari.23,mier.fetida.24}, we propose a setup with a chain of magnetic atoms arranged according to a one-dimensional (1D) Fibonacci quasicrystal \cite{jagannathan.21}.
Quasicrystals (QCs) \cite{shechtman.84,levine.steinhardt.84,levine.steinhardt.86,socolar.steinhardt.86,stadnik.98} have already been shown to host ordered states of matter including superconductivity~\cite{deguchi.15,fulga.16,sakai.17,kamiya.18,sakai.19,araujo.19,shiino.21}, while engineered quasiperiodicity~\cite{macia.06} has theoretically been predicted to enhance features such as superconducting proximity effect~\cite{rai.19,rai.20}, superconducting transition temperature and pairing amplitude~\cite{fan.21,zhang.22,sun.24,wang.24}, as well as the Josephson effect~\cite{sandberg.24}. 
In this work, we demonstrate how utilizing quasiperiodicity in an experimentally accessible setup generates an intriguing interplay between quasiperiodicity and topological superconductivity with MBS, leading to a substantially enlarged topological superconducting domain with increased stability of the MBS.

The defining feature of QCs is their lack of translational invariance, while still exhibiting long-range order through discrete scale invariance~\cite{janot.97}, which together with a non-crystallographic rotation symmetry yields well-defined diffraction peaks~\cite{hiller.85,goldman.93,baake.grimm.2012}.
This peculiar combination leads to fascinating properties, such as omnipresent criticality and multifractality~\cite{kalugin.katz.14,mace.jagannathan.piechon.16,mace.17}, as well as multiple topological invariants~\cite{fan.huang.21} that are otherwise only available in higher-dimensional periodic systems~\cite{kraus.13,flicker.wezel.15,kraus.16,huang.liu.18,petrides.18,huang.liu.19,chen.20,yamamoto.22}.
Significant advances in material control and fabrication have recently also led to improved measurement and growth techniques for QCs~\cite{nagao.15,wolf.21}, as well as to artificially engineered quasiperodic structures in reduced dimensions~\cite{guidoni.97,ledieu.04,sharma.05,ledieu.08,smerdon.08,talapin.09,lubin.12,forster.13,jia.16,bandres.16,collins.17,joon-ahn.18,yan.19,kuster.21,mahmood.21,xin.22,dang.23,uri.23}. % where there already exists a firm understanding of the fundamental QC topology, especially in 1D~\cite{jagannathan.21,fan.huang.21}. 
Motivated by these advances, we consider the 1D Fibonacci chain (FC)~\cite{jagannathan.21}, as it is one of the most theoretically studied~\cite{ostlund.83,kohmoto.83,ostlund.pandit.84,kohmoto.oono.84,tang.kohmoto.86,kalugin.kitaev.levitov.86,evangelou.87,kohmoto.sutherland.tang.87,sutherland.kohmoto.87,luck.89} and experimentally realized~\cite{steurer.sutter.07,negro.03,kraus.12,verbin.13,tanese.14,verbin.15,singh.15,dareau.17,baboux.17,lisiecki.19,goblot.20} quasiperiodic systems, known to host topological QC gaps with subgap winding states~\cite{kraus.12,kraus.zilberberg.12,verbin.13,tanese.14,verbin.15,rai.21,sykes.barnett.22} associated with a gap labeling theorem~\cite{bellissard.89,bellissard.92,mace.jagannathan.piechon.17}.

\begin{figure}
	\centering
\includegraphics[width=\linewidth]{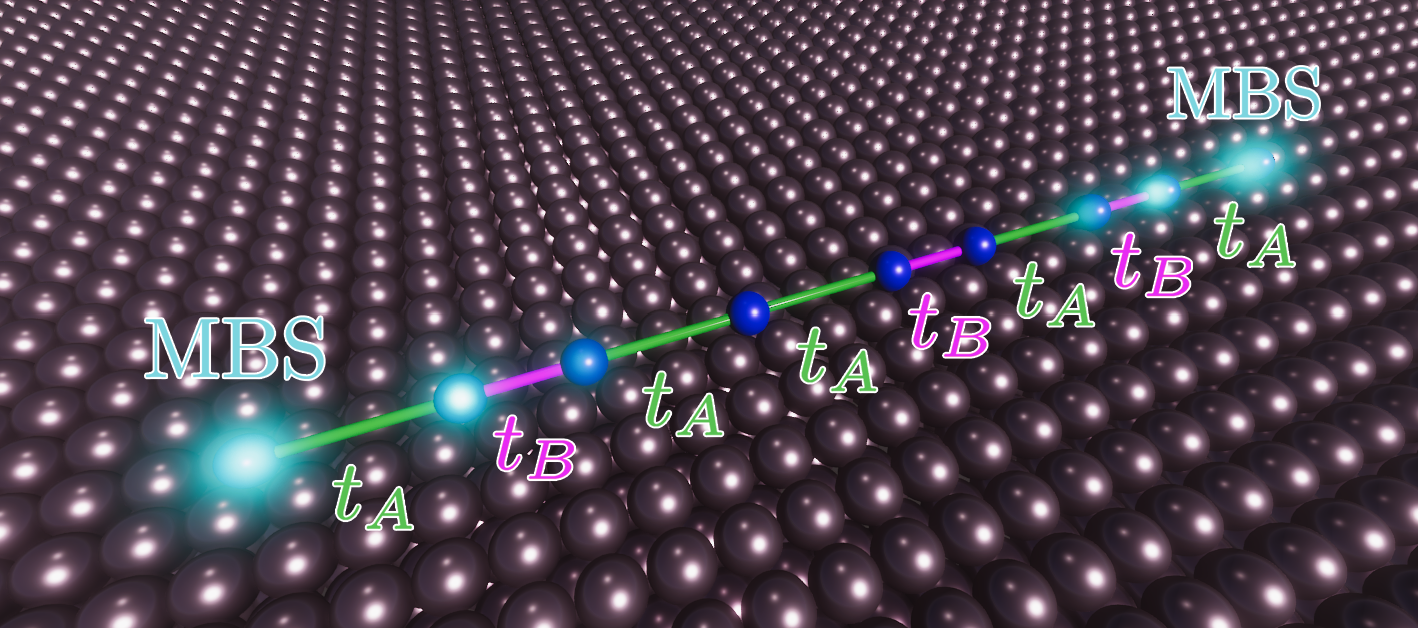}
	\caption{Schematic picture of the studied system showing a magnetic atomic chain of the $C_6$ Fibonacci approximant (blue) deposited on a superconducting surface (gray). Green and violet links represent the $t_A$ and $t_B$ hopping parameters, respectively. Teal-colored flares represent the MBS induced at the ends of the atomic chain in the topological phase. }
	\label{fig:Schematic}
\end{figure}
 
We note that some limited studies of MBS in quasiperiodic systems already exist in the context of the idealized spinless $p$-wave superconductivity in the Kitaev chain~\cite{ghadimi.17,jeon.22} or in the Aubry-Andr\'e-Harper (AAH) model~\cite{yahyavi.19}, as well as in non-Hermitian systems~\cite{liu.21} and in 2D~\cite{varjas.19,ghadimi.21,hori.sugimoto.24.arxiv}. For example, in the Kitaev model, Ref.~\cite{ghadimi.17} has found an onsite quasiperiodic modulation yielding a fractal structure on the boundary of the topological phase diagram, while Ref.~\cite{jeon.22} has found that the MBS may become more robust when the superconductor is quasiperiodic. 
Here, we both target an experimentally accessible setup and explicitly uncover the interplay of MBS and quasiperiodicity, with a resulting large impact on the overall phase diagram.

In particular, in this work we study an experimentally realizable arrangement of magnetic atoms in a chain following the 1D Fibonacci sequence and deposited on a superconducting substrate with intrinsic SOC. 
We find a substantially increased region of topological superconductivity with MBS in parameter space, while we at the same time uncover an intricate interplay between topological superconductivity hosting MBS and the topological properties of the QC. 
Overall, we find a staunch competition between a topological phase transition into the topological phase with MBS in the gap versus into a QC gapped phase, where only one of the two phenomenon survives beyond the transition. As a consequence, we find that the phase with MBS shows no sign of criticality or subgap winding otherwise found in QCs, and likewise, we find no MBS inside the QC gaps.
By investigating the behavior of the system as a function of magnetic field strength, chemical potential, and the quasiperiodic degrees of freedom, we are able to directly quantify this competition, both in the parameters influencing the MBS formation and the quasiperiodic degrees of freedom. 
Despite the high degree of possibilities, we find in the end a simple and general criterion for which phase is realized, based only on the energy gap size. 
Although this competition may seem unfavorable for MBS, we surprisingly find that the presence of quasiperiodicity is actually favorable for MBS realization. This is because each QC gap closing (of which there are many) can lead to additional topological MBS phases, in what are otherwise trivial regions in the crystalline limit. 
We also find that the topological gap that protects the MBS can be increased by quasiperiodicity. 
Thus, MBS can both easily form in entirely new physical parameter regimes of the underlying crystalline system and be more stable, which allows for a much broader use of materials.
Moreover, we find that shorter QC realizations, the so-called approximants~\cite{goldman.93}, are more favorable for MBS formation, which are also beneficial for experimental realization. 
Altogether, our results establish quasiperiodic systems as a very appealing platform for creating MBS and, more generally, for studying the fundamental interplay between quasiperiodicity and other topological phases or phase-coherent states of matter. 

The rest of this work is organized as follows.
First, we describe the model in Sec.~\ref{sec.model}, as well as review the key concepts of the FC and topological superconductivity with MBS. We then demonstrate the fundamental interplay and competition between topological superconductivity with MBS and quasiperiodicity in Sec.~\ref{sec.res.interplay}. In Sec.~\ref{sec.res.phase} we compute and interpret the resulting topological phase diagrams, along with analyzing spectral features. In Sec.~\ref{sec.res.soc} we analyze the additional effects due to a quasiperiodic SOC. Finally, we conclude our findings in Sec.~\ref{sec.sum}.

%%%%%%%%%%%%%%%%%%%%%%%%%%%%%%%%%%%%%%%%%%%%%%%%%%
\section{Model and background}
\label{sec.model}
%%%%%%%%%%%%%%%%%%%%%%%%%%%%%%%%%%%%%%%%%%%%%%%%%%
We start by specifying our theoretical model to study the interplay between topological superconductivity hosting MBS and quasiperiodicity. 
In particular, we consider the tight-binding Hamiltonian $H$ describing a 1D atomic chain (schematically shown in Fig.~\ref{fig:Schematic}):
\begin{equation}\label{eq:TotalH}
H = H_{\rm FC} + H_{\rm R}+ H_{\Delta}.
\end{equation}
Below we first focus on the term $H_{\rm FC}$, which models a FC consisting of magnetic atoms and review its properties, especially focusing on the important QC gaps and their topological properties.
Then, in order to generate topological superconductivity, the chain is assumed to be deposited on a superconducting substrate with SOC, which adds the proximity-induced Rasbha SOC $H_{\rm R}$  and superconductivity $H_{\Delta}$ into the chain. We introduce these terms and review how to quantify the topological superconducting state and its MBS.

\subsection{Fibonacci chain}
\label{sec.model.qc}
Motivated by the experimental developments in scanning tunneling microscopy (STM) techniques fabricating atomic chains on substrates \cite{cyrus.lutz.06}, we define a 1D structure of magnetic atoms following the Fibonacci sequence.
The first term in $H$, Eq.~(\ref{eq:TotalH}), thus describes itinerant electrons of a magnetic chain as
 \begin{eqnarray}\label{eq:NormalH}
 H_{\rm FC} &=&  \sum_{i, s,\sigma} \left( \mu + V_{\rm Z}\sigma_{\sigma\sigma}^z \right) c_{i s \sigma}^\dagger c_{i s \sigma}^{} \\ \nonumber 
&+& \Bigg(
 \sum_{ i, \langle ss^\prime \rangle,  \sigma} t_{\langle ss^\prime \rangle}  c_{i s \sigma}^\dagger c_{i s^\prime \sigma}^{} 
+  t_{\langle ss^\prime \rangle}  c_{is\sigma}^\dagger c_{i+1 s^\prime\sigma} + {\rm H.c.} \Bigg),
 \end{eqnarray}
where the operator $c_{i s \sigma}^\dagger$ creates an electron of spin $\sigma$ on site $s$ of unit cell $i$ of the atomic chain. 
Here $\mu$ is the overall chemical potential, $V_{\rm Z}$ models the (classical) magnetic moment of each atom through an effective Zeeman splitting, and $t_{\langle ss^\prime \rangle}$ is the hopping between nearest neighbor atoms (equivalently sites), with $\langle \dots \rangle$ only selecting for nearest neighbor terms. 
In order to study QC-like structures, we consider a magnetic atomic chain composed of Fibonacci approximants $C_n$, which redesign the usual crystalline hopping structure by denoting them with indices $s\in [1,F_n]$, where $F_n$ is the $n$th Fibonacci number.
In order to form longer chains, the approximant, viewed as the QC unit cell, is then repeated such that $i\in \left[1,R\right]$, to guarantee a sufficiently long chain capable of hosting well-isolated MBS as its topological boundary states.
The full length of the chain is then given by $\Omega = R \times F_n +1$, with the extra site added to avoid dangling bonds.

Within each Fibonacci approximant, we use the Fibonacci hopping model~\cite{jagannathan.21} where the nearest neighbor hopping parameter takes one of two values, $t_{\langle ss^\prime\rangle }\in\{t_{\rm A},t_{\rm B}\}$. 
This connects the sites through two different bond types (sometimes referred to as atomic and molecular bonds depending on their relative strengths), which are determined according to a Fibonacci sequence. 
Specifically, the construction of the $n$th Fibonacci approximant $C_{n}$ can be obtained by concatenation ($\oplus$) via the recursion relation
\begin{equation}
\begin{split}
& C_1=t_B, \ C_2=t_A, \ C_3 = t_At_B, \ C_4 = t_At_Bt_A \ \text{etc.}\\
& C_n=C_{n-1} \oplus C_{n-2}, \quad (n > 2),\\
& F_n = {\rm length}[C_n] = {\rm length}[C_{n-1}] + {\rm length}[C_{n-2}],
\end{split}
\end{equation}
Thus systems with the unit cell described by $C_1$ or $C_2$ corresponds to a periodic chain with hopping parameters $t_B$ or $t_A$, respectively, while a unit cell described by $C_3$ corresponds to a dimerized model of repeating $t_At_B$ couplings, i.e.~akin to the Su-Schrieffer-Heeger (SSH) model~\cite{su.shrieffer.79,waka.ezawa.14,ezawa.17,hua.chen.19, kobialka.sedlmayr.19}. 
Larger $n$ give rise to increasing sizes of the approximants, see Table~\ref{tab:FibSeqSites}, which provides the details on the system sizes considered in this work. 
We choose the number of repetitions $R$ such that the full size of the system $\Omega$ is as close to $300$ sites as possible for all studied FCs. 
This ensures that the nanowire is sufficiently long to observe isolated MBS behavior.

We note that a more generalized Fibonacci approximant can be obtained by instead assigning the nearest neighbor hopping parameters via the characteristic function~\cite{kraus.zilberberg.12}
\begin{equation}
    \label{eq:phason}
    \langle s s^\prime \rangle  = \sgn \left[\cos{\left( \frac{2  s \pi }{\tau}  -\frac{\pi }{\tau} + \phi \right)} - \cos{\left(\frac{\pi }{\tau}\right)}\right],
\end{equation}
where a positive (negative) value is substituted with the hopping $t_A$ ($t_B$), and where $\tau = (1+\sqrt{5})/2$ is the golden ratio, while the phase factor $\phi \in [0,2\pi)$ is called the phason angle.
Varying the phason angle thus flips certain hoppings in a specific manner, creating a family of $F_n+1$ unique realizations of the $C_n$ Fibonacci approximant~\cite{jagannathan.21}.
From now on, we express all energies in units of $t_A$ and introduce the hopping ratio $\rho \equiv t_B/t_A$, such that $\rho = 1$ and $\rho \neq 1$ describe crystalline and quasiperiodic systems, respectively.

\begin{table}[t]\centering
    \begin{tabular}{ |c|c|c|c|c|c|c|c|c|c|c|c|c| }
        $C_n$ & $C_2$ & $C_3$ & $C_4$ & $C_5$ & $C_6$ &  $C_8$ &  $C_{11}$ & $C_{13}$	\\
        \hline
        $F_n$ & 1 & 2 & 3 & 5 & 8 & 21 &   89 &  233\\
        $R$ & 300 & 150 & 100 & 60 & 38 &  14 &  3 &  1
    \end{tabular}
    \caption{Fibonacci approximant s$C_n$ with the number of bonds given by the corresponding Fibonacci number $F_n$, while $R$ corresponds to the number of repetitions of $C_n$ (i.e.~for $R>1$, $C_n$ gives the unit cell structure). 
    The total number of sites in the chain is $\Omega=R\times F_n+1$.
    }
    \label{tab:FibSeqSites}
\end{table}

\begin{figure*}
\includegraphics[width=\linewidth]{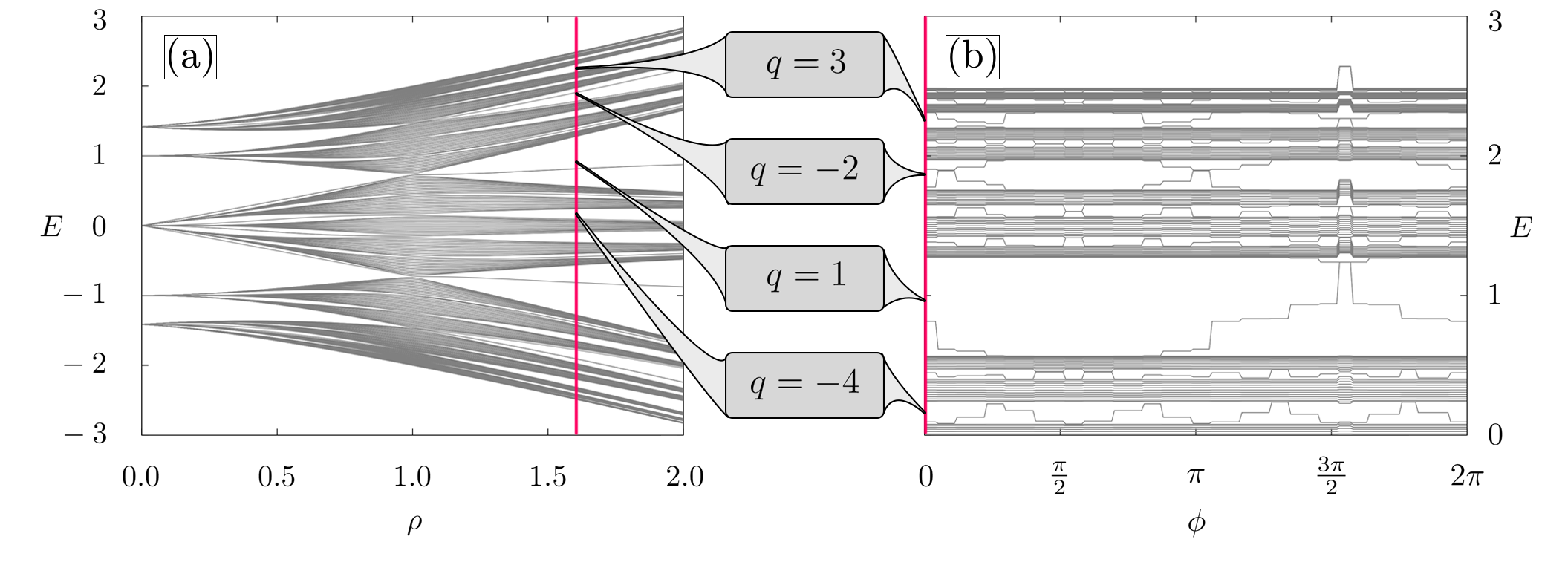}
\caption{(a) QC energy spectrum versus hopping ratio $\rho$ in a Fibonacci atomic chain with $C_8$ (21 bonds) repeated $R=14$ times ($\Omega=295$ sites), using $\mu=\BZ=\Delta=\alpha=0$ and $\phi=0$.
(b) QC energy spectrum for the same approximant but as a function of the phason angle $\phi$ at fixed $\rho=1.6$ [red line in (a)], illustrating that each QC gap has a topologically protected subgap state winding across the gap as a function of $\phi$ according to the gap label $q$ provided by Eq.~(\ref{eq:gap_labeling_theorem}).}
\label{fig:QC_gaps}
\end{figure*}

\subsection{Quasicrystal gap labeling}
\label{subsec.GCgap}
In order to properly understand the possible emergence of topological superconductivity in the FC, we first need to take into account the already inherent QC topology.
Below, we review the existence of topological QC gaps and their gap labeling theorem in the FC, as well as the related subgap states and their winding across the gaps with the phason angle.

In the FC, $\rho \neq 1$ leads to the opening of a multitude of energy gaps in the spectrum, which are quantified by a gap label $q$~\cite{bellissard.89,bellissard.92,mace.jagannathan.piechon.17}, see Fig.~\ref{fig:QC_gaps}. 
Within each gap there are also subgap states that typically have a critical and multifractal behavior~\cite{jagannathan.21}.  
In fact, it has recently been shown that the 1D FC is topologically equivalent to both the 2D quantum Hall and the 1D AAH models~\cite{kraus.zilberberg.12,flicker.wezel.15,kraus.16}, and therefore the gap label $q$ can be identified as a Chern number, which also describes the winding of the subgap states across the gap~\cite{kraus.12}.
However, instead of winding in the traditional sense (i.e.~in momentum space), the states wind $|q|$ times across the gap with direction $\sgn(q)$ as a function of the phason angle $\phi$, as illustrated in Fig.~\ref{fig:QC_gaps}(b).
Interestingly, the number of gaps and the value of $q$ for approximant $C_n$ follow a gap labeling theorem~\cite{bellissard.89,bellissard.92,mace.jagannathan.piechon.17},
\begin{equation}
    \label{eq:gap_labeling_theorem}
     \frac{\mathcal{N}_E}{4\Omega }
    = \operatorname{mod}\left(\frac{qF_{n-1}}{F_n},1\right),
\end{equation}
where $\mathcal{N}_E$ is the number of bands below energy $E$.
Thus, by calculating $\mathcal{N}_E$, the gap labeling theorem can be used to extract the gap label $q$.

It has recently been shown that the gap labeling theorem also provides additional information for approximants $C_n$, specifically the existence of stable (transient) gaps that remain (disappear) for infinitely long QCs~\cite{mace.jagannathan.piechon.17}.
Here, the largest gaps are generally stable and also associated with the smallest $q$, which also turn out to be the most relevant gaps for our work. 
In terms of the subgap states, they do not only wind across the energy gap $|q|$ times as a function of the phason angle, but also their real space position varies along the FC $|q|$ times~\cite{jagannathan.21}. As a result, for certain phason angles and thus certain realizations of the Fibonacci approximant, these subgap states may also appear as (real-space) edge states. 
Figure~\ref{fig:QC_gaps} further shows that multiple subgap states can also exist in some gaps.
Notably, all these properties of the FC clearly go beyond the SSH chain, where only a single energy gap opens with a single set of edge modes appearing in the topological regime ($\rho <1$)~\cite{su.shrieffer.79,waka.ezawa.14,ezawa.17,hua.chen.19, kobialka.sedlmayr.19}.

\subsection{Superconducting surface}
\label{sec.model.tsc}
It is well-known that a crystalline magnetic atom chain deposited on a superconducting surface with SOC, equivalently described by Eq.~\eqref{eq:TotalH}, can host topological superconductivity \cite{lutchyn.sau.10,oreg.refael.10,nadjperge.drozdov.13,nadjperge.drozdov.14}.
In order to investigate the impact of quasiperiodicity, we similarly define the terms describing the effect of the superconducting substrate on the atomic magnetic chain. 
Here, it is the combination of proximity-induced SOC $H_{\rm R}$ and superconductivity $H_{\Delta}$ in Eq.~(\ref{eq:TotalH}) into the chain that is crucial for topological superconductivity with MBS.

The SOC, modeled by $H_{\rm R}$, describes an effective Rashba SOC, which can be strong either in heavy elemental superconductors, such as Pb~\cite{nadjperge.drozdov.14}, or alternatively arise from a noncollinear magnetic ordering of the atoms in the chain, such as for an Fe chain on a Re surface~\cite{kim.palaciomorales.18}) and is given by
\begin{align}
\label{eq:HR}
H_{\rm R} = \sum_{ i, \langle ss^\prime \rangle,  \sigma, \sigma^\prime} c_{is\sigma}^\dagger \left[\alpha \sigma_{\sigma\sigma^\prime}^y \right] \left(  c_{i s^\prime \sigma'}^{}
+ c_{i+1 s^\prime\sigma'}\right) + {\rm H.c.} 
\end{align}
with $\alpha$ the SOC strength and $\sigma_{\sigma \sigma'}^{\nu}$ denoting the Pauli matrices, $\nu \in \{x,y,z\}$. 
We model the superconducting pairing induced into the atomic chain through
 \begin{equation}\label{eq:HDelta}
 H_{\rm \Delta} =  \sum_{i, s} \Delta c_{i s \uparrow}^\dagger  c^\dagger_{i s \downarrow} + {\rm H.c.},
\end{equation}
where $\Delta$ is the effective superconducting order parameter, here chosen to be a real value without loss of generality.
We here assume conventional $s$-wave spin-singlet superconductivity, which is also constant along the whole chain, modeling a large substrate with small effect of the magnetic atoms.

We diagonalize the resulting Hamiltonian $H$ in Eq.~(\ref{eq:TotalH}) using the Bogoliubov-de-Gennes formalism at zero temperature, using open boundary conditions.
We vary most of the physical parameters in the model to capture the full interplay between topological superconductivity and quasiperiodicity.
For illustrative purposes, we often only vary one or two parameters at a time, while keeping the others constant.
Typical constant values used are (unless otherwise specified) $\mu=-2 t_A$, $\Delta=0.2 t_A$, $\alpha=0.2 t_A$, $V_Z=0.4 t_A$, using $t_A$ as a unit of energy. 
These choices generally ensure the emergence of MBS in a typical crystalline atomic chain.

\subsection{Topological classification}
\label{sec:topoclass}
Having defined the model, we next elucidate how to quantify its topological properties. 
We already discussed the topological properties of the bare FC in Sec.~\ref{subsec.GCgap}, including the gap labeling theorem and its associated winding of the subgap states with phason angle. 
Adding superconductivity and SOC to the chain also generates the possibility of a topological superconducting phase with MBS.
A common and straightforward way of studying the emergence of MBS is to compute the topological phase transition and an appropriate topological invariant within its BDI topological symmetry class \cite{tewari.sau.12, fidkowski.kitaev.11, kobialka.sedlmayr.19,kobialka.sedlmayr.20}.
Unfortunately, a quasiperiodic system cannot be described in reciprocal space as it lacks spatial periodicity, similarly to amorphous materials and other non-periodic systems. 
As standard topological invariants are defined in reciprocal space, such an approach is thus unavailable. Alternatively, a few real-space approaches have been developed to describe topology in real-space systems, e.g.~based on local Chern markers~\cite{kitaev.06,prodan.10,flicker.wezel.15,marsal.varjas.20,hannukainen.martinez.22,corbae.hannukainen.23}, the Bott index~\cite{loring.hastings.11}, the Majorana polarization (MP)~\cite{sticlet.bena.12,sedlmayr.bena2015,awoga2024majorana}, or relying on projections from a higher-dimensional periodic system via hidden dimensions \cite{jagannathan.21,fan.huang.21}. 
To be able to effectively distinguish the different types of topologies, we here focus on a hybrid approach, which we find suitable for large-scale numerical screening of a large parameter space. It uses several ingredients: the MP, the standard deviation of energy levels with varying phason angle $\phi$, and tracking the origins of the energy gaps near zero energy. 
We also complement and check our results using a Wilson loop real-space method~\cite{Wang2019Band,Awoga2022Robust}, as described in Appendix~\ref{app.res.wilson}, and find excellent agreement. This fully validates the finding of topological superconductivity with MBS.

The MP quantifies to what degree zero-energy states possess MBS character and we base its form on~\cite{sticlet.bena.12,awoga2024majorana}
\begin{eqnarray}
        {\rm MP}_d &=& \frac{\sum_{i \in d} u_{in} v_{in}^{\ast}+ v_{ni} u_{in}^{\ast}}{\sum_{i \in d} |u_{in}|^2 + |v_{in}|^2}, \\
    {\rm MP}&=&{\rm MP}_{\rm L}^{} {\rm MP}_{\rm R}^{} ,
\end{eqnarray}
where $u_{i}$ and  $v_{i}$ are Bogoliubov-de Gennes eigenvectors for $n=2\Omega$ and $n=2 \Omega +1$ (i.e.~the states closest to zero energy), while ${\rm L}$ and ${\rm R}$ denote the left and right halves of the FC.
A value of the MP sufficiently close to $-1$ signifies that the system is capable of hosting MBS, while a positive value informs that the system does not have MBS. 
The latter would mean that the system either hosts the QC phase or a trivial superconducting (SC) phase near zero energy.
For regions in phase space where the MP shows the existence of MBS, we extract the size of the energy gap (from the eigenvalues of the system) to obtain the size of the topological gap $\DeltaT$. 
This gap determines the robustness of the topological phase as it is the gap protecting the MBS from all other quasiparticles.
It is also valuable to note that this method can be experimentally validated using spin-polarized Andreev spectroscopy~\cite{he.ng.14,wang.geng.23}.
For ease of communication, we occasionally call the topological phase with MBS simply the topological nontrivial phase, and refer to the other phases then as topologically trivial, although we note that these can still host QC topology~\cite{jagannathan.21}. 

To distinguish QC topology from the truly topologically trivial superconducting state (SC), we use the existence of a subgap state that winds with phason angle, which is a key property of QC topology. 
In this context, it is important to vary the phason angle to track the winding, as for certain values (and thus a certain approximant) the QC subgap state might lie at the edge of the gap, resulting in a false positive indication of a full trivial SC gap.
Thus, in order to efficiently distinguish the QC phase, we extract the standard deviation of the energy levels $E_{2\Omega+1}$ (can be subgap state or MBS) and $E_{2\Omega+2}$ (can be subgap or state at the band edge) when varying the phason angle $\phi \in [0,2\pi)$.
We find that a standard deviation larger than $10^{-3} t_A$ for any of these states is sufficiently large to clearly indicate the existence of subgap states, thus marking the QC phase. 
We henceforth refer to the gap within the QC topological gap as $\DeltaQC$.
We complement using the MP and the spread in energy with phason angle to find the topological MBS and topological QC phases, respectively, by also tracking their energy gaps, $\DeltaT$ and $\DeltaQC$, from regions in parameter space where we already know the nature of the phase. 
We further complement these results with using a Wilson loop real-space method~\cite{Wang2019Band,Awoga2022Robust} (see Appendix~\ref{app.res.wilson}).
We find that all these methods show excellent agreement. 
Finally, we refer to regions in parameter space without MBS and no phason angle winding of subgap states, as a trivial superconducting (SC) phase, with gap $\DeltaSC$.

\begin{figure*}[t]
\includegraphics[width=\textwidth]{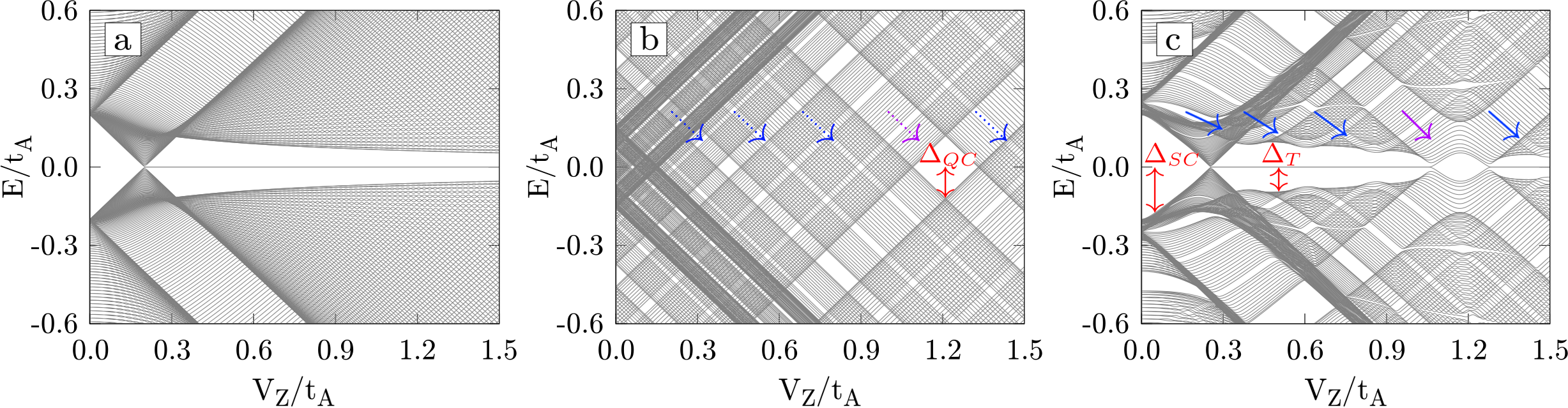}
\caption{Energy spectra versus magnetic field $V_{\rm Z}$ for the approximant $C_{8}$ ($R=14$ repetitions) at $\mu=-2t_A$. (a) Crystalline atomic chain ($\rho=1$) with MBS beyond the topological phase transition ($\Delta=0.2t_A, \alpha=0.2t_A$). (b) QC chain ($\rho=1.2$) without MBS ($\Delta=\alpha=0$). (c) QC chain ($\rho=1.2$) with MBS ($\Delta=0.2t_A, \alpha=0.2t_A$). Dotted arrows indicate QC gaps, blue arrowed gaps are ``reflected" against the dominating MBS gap and thus do not survive at zero energy, violet arrowed gap is larger than the MBS gap and instead overtakes the MBS gap at zero energy.
Labels denote trivial superconducting gap $\DeltaSC$, topologically nontrivial MBS gap $\DeltaT$, and QC energy gap $\DeltaQC$.
} 
\label{fig:Spectra}
\end{figure*}
%%%%%%%%%%%%%%%%%%%%%%%%%%%%%%%%%%%%%%%%%%%%%%%%%%
\section{Interplay between topological superconductivity and quasiperiodicity}
\label{sec.res.interplay}
To start our study of topological superconductivity in a QC system, we first explore the energy spectrum, aiming to elucidate the gap structure of both topological superconductivity and quasiperiodicity. 
For pedagogical purposes, we first consider these phenomena separately [Fig.~\ref{fig:Spectra}(a) and (b), respectively], and then combine them to highlight their interplay  [Fig.~\ref{fig:Spectra}(c)]. 
Finally, last in this section, we confirm the interpretations by also displaying the phason angle dependence.

Figure~\ref{fig:Spectra}(a) shows the atomic chain without quasiperiodicity ($\rho=1$ or equivalently the $C_1$ approximant), which exhibits a topological phase transition with an accompanied bulk gap closing when tuning $V_z$. 
The emergence of nontrivial topology with MBS takes place for magnetic field $\Bcp < \BZ < \Bcm$ given by
\begin{equation}
    \label{eq:critical_B}
    V_\pm^{C_1} = \sqrt{(\mu \pm 2 t_B)^2 + \Delta^2}.
\end{equation}
With our choice of parameters $\Bcp = 0.2t_A$ in Fig.~\ref{fig:Spectra}(a) and $\Bcm\approx4.0t_A$ (not shown).
The lower critical field corresponds to the closing of the spin-singlet $s$-wave gap due to the Zeeman spin splitting reaching the level of the gap itself above which the SOC enables the opening of an effective spinless $p$-wave superconducting gap.

Next, Fig.~\ref{fig:Spectra}(b) shows a FC ($\rho=1.2$) as a function of magnetic field, but without superconductivity and SOC such that there cannot be any MBS present.
At $\BZ=0$ the spectrum is spin degenerate with multiple QC gaps according to the gap labeling theorem, Eq.~(\ref{eq:gap_labeling_theorem}), and, analogously to Fig.~\ref{fig:QC_gaps}(a), here with the largest gaps at energies $E \approx 0.45t_A$ and $E \approx 1.2t_A$ (the latter is not shown at $\BZ=0$). 
As $\BZ$ increases, the energy levels spin split linearly, such that these largest QC gaps occur at zero energy at $\BZ \approx 0.45 t_A$ and $\BZ \approx 1.2 t_A$, overall leading to multiple QC gap openings at zero energy when tuning $\BZ$, see the dotted arrows indicating the evolution of gaps with field.
We note that varying the chemical potential $\mu$ instead (as later in Sec.~\ref{sec.res.phase_diagram}) leads to an analogous scenario where the QC gaps also occur at zero energy, i.e.~the Fermi level is tuned to the QC gaps. 

Finally, Fig.~\ref{fig:Spectra}(c) shows the combined influence of topological superconductivity and quasiperiodicity.
Here we still observe a topological phase transition into a phase with MBS, but now for a slightly larger $\Bcp \approx 0.28t_A$. 
This is due to the larger $t_B$, not captured by Eq.~(\ref{eq:critical_B}).
Later, in Sec.~\ref{sec.res.phase}, we demonstrate how the analytic expression for $\Bcpm$ can be generalized with the corresponding quasiperiodic structure, yielding the same transition as in these numerical results.
The figure further shows that as $\BZ$ increases, multiple QC gaps ($\DeltaQC$) move down toward zero energy (arrows), just as in Fig.~\ref{fig:Spectra}(b). 
However, we find that the smaller QC gaps never reach zero energy, but they are instead ``reflected'' against the MBS gap, back into the continuum (blue arrows). 
Each such QC gap reflection leads to a small decrease of $\DeltaT$ at the corresponding value of $\BZ$. 
A cardinal, but opposite, example of this antagonistic behavior is found around $\BZ=1.2t_A$ (violet arrow). 
Here $\DeltaQC > \DeltaT$, resulting instead in the closing of the topological gap and a QC gap is opened at zero energy, with the consequence that the MBS disappears at a phase transition into the QC phase, but then reappears again as $\DeltaQC$ closes at an even higher $\BZ$.
Explained from another perspective, we find that superconductivity gaps out the spectrum, thus shifting all energy levels and QC gaps to the continuum as long as $\DeltaQC < \DeltaT$. 
However, when $\DeltaQC > \DeltaT$, there are no energy levels below $\DeltaSC$ to participate in the condensation, and the QC gap survives.
As a consequence, the MBS and QC gap structures and topologies are competing for the Fermi surface instability of the system and are hence mutually exclusive of each other.

\subsection{Phason angle dependence}
\label{sec:phason}
In order to further cement the antagonistic behavior of the MBS and QC phases found in Fig.~\ref{fig:Spectra}, we next study their behavior as a function of phason angle. 
This has the added benefit of showing how using the variation in energies of the subgap states with phason angle is an effective tool to characterize the QC phase, as described in Sec.~\ref{sec:topoclass}. 

\begin{figure}[b]
	\centering
 	\includegraphics[width=1\linewidth]{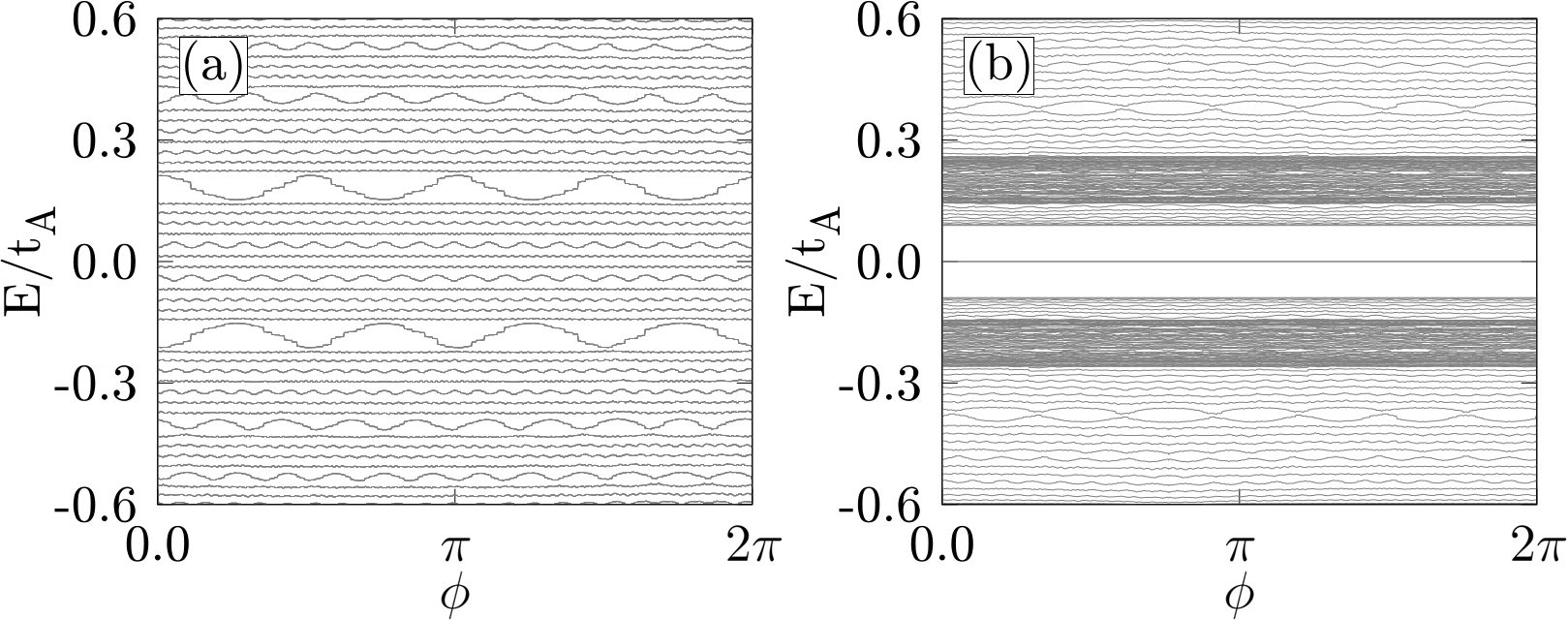}
	\caption{Energy spectrum of the Fibonacci approximant $C_{13}$ versus phason angle $\phi$ for (a) bare FC ($\mu=\Delta=\alpha=\BZ=0t_A$, $\rho = 1.2$) and 
(b) example of superconducting FC typically studied in this work ($\mu=-2.2t_A$, $\Delta=0.2t_A$, $\alpha=0.2t_A$, $V_Z=0.4t_A$, $\rho=1.2$).  
 }
	\label{fig:phason}
\end{figure}

In total, we need to distinguish between three possible different phases in the system:
a topological phase capable of hosting MBS, a QC phase, and a trivial gapped SC phase. 
Here, both of the first two phases contain subgap states. For the QC phase, the subgap state additionally winds around the gap with the phason angle $\phi$.
However, at certain values of $\phi$, the winding state may reside at the gap edge and thus the QC gap can easily mimic a fully gapped SC state at least for some phason angles.
Furthermore, it is a priori not clear if the MBS subgap state in the topological phase will also display phason angle winding if the system is a QC.
To clarify these questions we vary the phason angle $\phi$, which cycles through all possible combinations of $A$ and $B$-type hopping, composing the full family of each Fibonacci approximant. 
Figure~\ref{fig:phason} shows a comparison between two prototype realizations of the $C_{13}$ Fibonacci approximant as a function of the phason angle $\phi$; (a) a bare FC ($\mu=\Delta=\alpha=\BZ=0t_A$) and (b) the same FC but with typical parameters used in this work ($\mu=-2 t_A, \Delta=0.2 t_A, \alpha=0.3 t_A, \BZ=0.3 t_A$) to produce topological superconductivity.
For the bare FC (a), we find a clear winding of the subgap states in each QC gap. 
As such, extracting the standard deviation of the energy levels when they vary through the phason angle clearly offers a numerically robust characterization of the QC phase.
However, for the FC with superconductivity and SOC (b), the existence of superconductivity entirely removes the winding subgap states around zero-energy and instead MBS appear at zero energy, which are notably entirely unperturbed by the change in phason angle. Further, the size of the topological gap $\DeltaT$ is also not varying with the phason angle.
QC substates at higher energies, outside the topological gap $\DeltaT$, still winds, but across notably reduced QC gaps.
Furthermore, we do not find that the MBS exhibits any critical behavior typical for quasiperiodic states (not shown).
This demonstrates not only that topological superconductivity with MBS and QC topology with winding subgap states are antagonistic to each other, but also that their defining characteristics are mutually exclusive of each other.

For the case of the topological superconducting phase with MBS, we further find that varying $\phi$ does not influence the overall boundaries of the topological regions either. 
A minor exception exists only for a single element of each Fibonacci approximant family, e.g.~$\mathbf{t_Bt_At_B}$ for $C_4$ or $\mathbf{t_Bt_At_Bt_A}t_A\mathbf{t_Bt_At_B}$ for $C_6$, where bold letters signify the hopping opposite from the one expected in the typical corresponding approximant, i.e. at phason angle $\phi=0$. 
This exception arises due to the Fibonacci approximant gaining mirror symmetry, which can have a minor effect of the overall outline of the different phases. 
In the following, we avoid this special fine-tuned mirror-symmetric situation and at the same time always check the results for a multitude of phason angles, e.g.~using $7$ separate values for $C_{13}$, as we find that a further increase does not change the results quantitatively.

%%%%%%%%%%%%%%%%%%%%%%%%%%%%%%%%%%%%%%%%%%%%%%%%%%%
%%%%%%%%%%%%%%%%PHASE DIAGRAM%%%%%%%%%%%%%%%%%%%%%%
%%%%%%%%%%%%%%%%%%%%%%%%%%%%%%%%%%%%%%%%%%%%%%%%%%%
\section{Topological phase diagrams}
\label{sec.res.phase}

The results in Figs.~\ref{fig:Spectra} and \ref{fig:phason} highlight the fundamental and also mutually exclusive interplay between the energy and topology of the MBS and QC phases, which at a first glance seemingly suggests that the formation of a QC atomic chain is unfavorable to generate topological superconductivity with MBS. 
However, these results display only a single slice of a much richer topological phase diagram.
In this section, we quantify the interplay between the QC gap structure and MBS topology by calculating the full topological phase diagrams. 
Surprisingly, we find additional and large topological regions with MBS in the phase diagram that cannot be obtained for a typical crystalline atomic chain.
In order to show the full picture we first compute the topological phase diagrams as a function of the quasiperiodic degrees of freedom and chemical potential, and then also as a function of the magnetic field.

%%%%%%%%%%%%%%%%%%%%%%%%%%%%%%%%%%%%%%%%%%%%%%%%%%%
%%%%%%%%%%%%%%%%CHARACTERIZAION%%%%%%%%%%%%%%%%%%%%%%
%%%%%%%%%%%%%%%%%%%%%%%%%%%%%%%%%%%%%%%%%%%%%%%%%%%
\subsection{Hopping ratio $\rho$ dependence}
\label{sec.res.phase_diagram}

\begin{figure*}[t!]
\includegraphics[width=\linewidth]{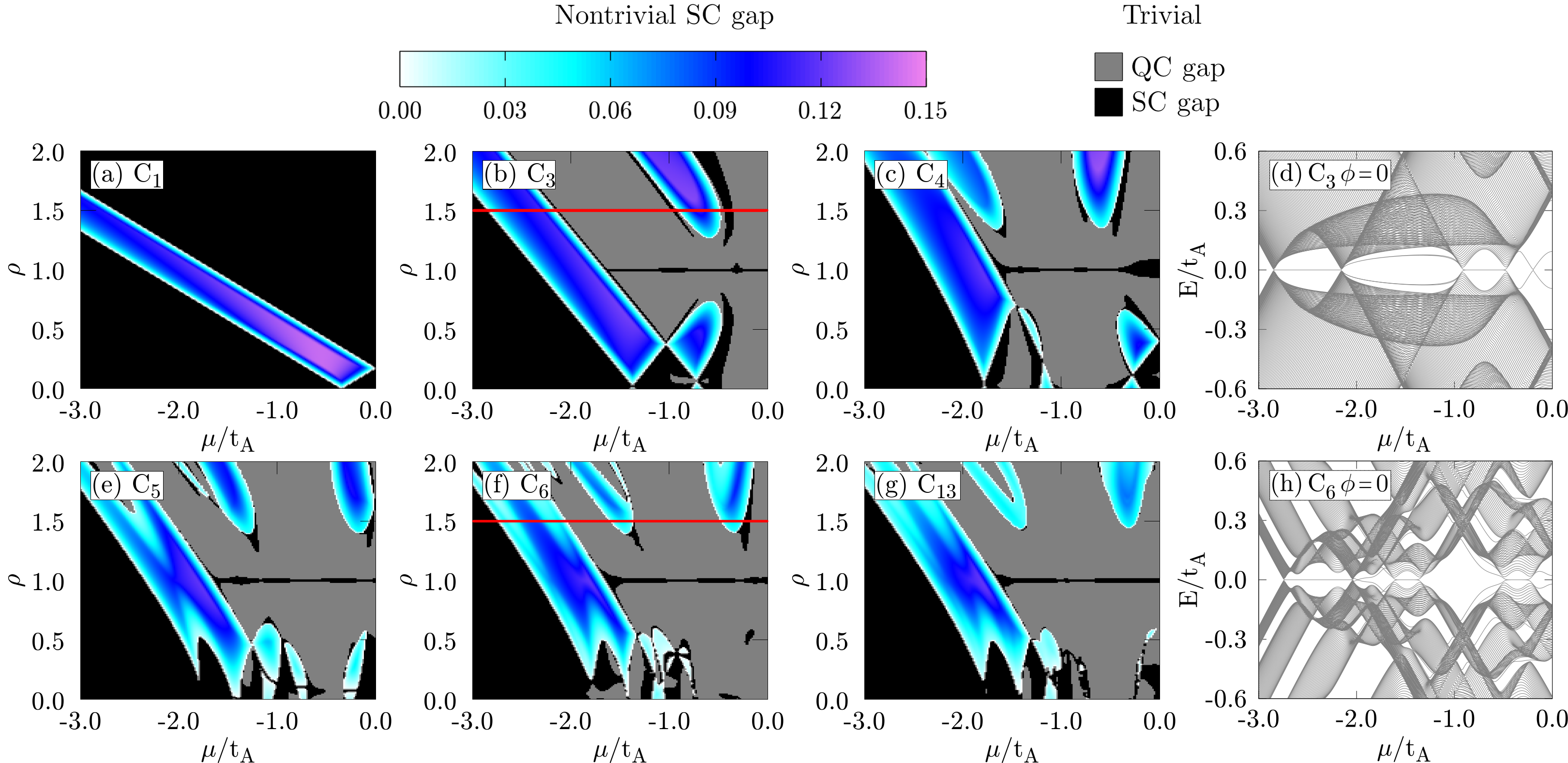}
\caption{
Phase diagrams as a function of chemical potential $\mu$ and hopping ratio $\rho$ obtained for $C_1$, $C_3$, and $C_4$ (top row),and  $C_5$, $C_6$, and $C_{13}$ (bottom row) approximants.
Panels (d,h) show the energy spectrum as a function of $\mu$ for $C_3$ and $C_6$ at the red lines in (b,f), respectively. 
Parameters used: $\BZ=0.4t_A$, $\Delta=0.2t_A$, $\alpha=0.2t_A$. 
Panels (a,d,h) have $\phi=0$, while the rest of the panels are generated for all relevant phason angles $\phi$ that force the QC subgap state to wind around the gap.
} 
\label{fig:map_mr}
\end{figure*}

We start by studying the topological phase diagram as a function of the hopping ratio $\rho$ and the chemical potential $\mu$ for different Fibonacci approximants $C_n$ in Fig.~\ref{fig:map_mr}. 
Here we display regions with MBS with its corresponding gap size in color.
To avoid confusion, we refrain from showing the size of the QC (grey) and SC (black) gaps, hence the discrete scale for these phases.
In Fig.~\ref{fig:map_mr}(a) we plot the simplest approximant, $C_1$, a chain composed only of $t_B$ hoppings, thus fully crystalline. 
Increasing $\rho$ in this case only makes the chain squeezed along its axis in real space.
The nontrivial topological phase emerging in the phase diagram thus shows the linear dependence between $\mu$ and $t_B$, as expressed by Eq.~(\ref{eq:critical_B}). 
Moreover, the largest topological gap, protecting the MBS, is observed in the region of a small hopping ratio and $|\mu|\approx t_B \approx \Delta$ (violet). 
The next approximant, $C_2$, consists of only $t_A$ hoppings and is thus also crystalline, and we therefore refrain from showing any $C_2$ results.

In Fig.~\ref{fig:map_mr}(b) we plot the topological phase diagram for the $C_3$ approximant. 
Here, the inclination of the main topological phase changes, but it still remains linear.
Most interestingly, a new set of {\it satellite} topological phases appears, except near $\rho \approx 1$ where the system is close to the crystalline chain.
These satellite topological phases are regions in parameter space, where quasiperiodicity allows for topological superconductivity with MBS outside of parameter regions available for the crystalline chain.
Interestingly, the topological phase emerging in the high $\rho$ region possesses the largest topological gap, showing enhanced robustness of the topological phase.
As the $C_3$ chain is similar to the SSH model, we infer that the emergence of satellite topological phases can be seen as the result of the odd number of parity inversions of the negative energy bands~\cite{kobialka.sedlmayr.19}.
In Fig.~\ref{fig:map_mr}(d) we plot the energy spectrum along the red line in (b). 
Here, the MBS in both the original and satellite regions are clearly visible. 
In between these, two bulk gap closings near $\mu\approx-2.2t_A$ and $-1.0t_A$ mark transitions between the MBS and QC phases, where in the QC phase a clear subgap state is present, which also changes its energy if the phason angle is varied (not shown). 
Notably, even this short approximant displays clear QC behavior, which fills essentially the same overall part of the phase diagram as for all non-crystalline approximants (beyond new satellite topological MBS regions). 

Interestingly, in the close vicinity of the topological regions we usually find small slivers of a trivial SC gap (black) for all approximants repeated sufficiently to create long chains. 
This creates a buffer zone between the MBS and QC phases, where first the bulk gap closes, marking the end point of the MBS phase (color), but with QC subgap states (grey) only appearing once a sufficient bulk gap is formed.
This also implies that smaller Fibonacci approximants must be adequately repeated ($R>1$) to be useful for MBS formation or no topological phase with a sufficiently large energy gap can emerge, equivalently to standard crystalline chains where sufficient length is required to separate the two MBS at the chain end points.
In Appendix~\ref{app.repetition} we provide additional data on the $R$ dependence and conclude that for sufficiently long chains the outline of the topological phase is independent of the number of repetitions of the Fibonacci QC.

We next consider $C_4$ in Fig.~\ref{fig:map_mr}(c), which contains an additional bond compared to $C_3$, and we also find richer QC features. 
In this case, the original topological phase bends somewhat, losing its simple linear dependence of its boundaries. 
Additional satellite nontrivial MBS phases also appear, overall covering a further increasing region in parameter space compared to the crystalline chain.
Similarly to $C_3$, the satellite MBS region at high value of $\rho$ features the largest topological gap.
We find that this enhancement of the topological gap in satellite regions due to the QC structure ceases for $C_5$, see Fig.~\ref{fig:map_mr}(e).

Moving to higher Fibonacci approximants shows emergence of yet additional satellite topological MBS phases, see Fig.~\ref{fig:map_mr}(e-g). 
However, an increased number of overlapping nontrivial phases leads to gap closings and fragmentation of these nontrivial regions, resulting in either the complete removal of the nontrivial MBS phase or lifting of the MBS from zero energy due to too small protecting energy gaps. 
This can especially be seen at the extremes of the $\rho$ parameter range and becomes more severe for the higher order approximants. 
As an example, Fig.~\ref{fig:map_mr}(h) depicts the energy spectrum along the red line in (f).
We observe multiple closings and reopenings of the energy gap, where the trivial gaps are either of MBS or QC nature.
Finally, we note that we find qualitatively similar phase diagrams when considering different models for the hopping ratio, e.g.~with conserved versus varying bandwidth, see Appendix~\ref{app.bandwidth}.

\begin{figure*}[t]
\includegraphics[width=\linewidth]{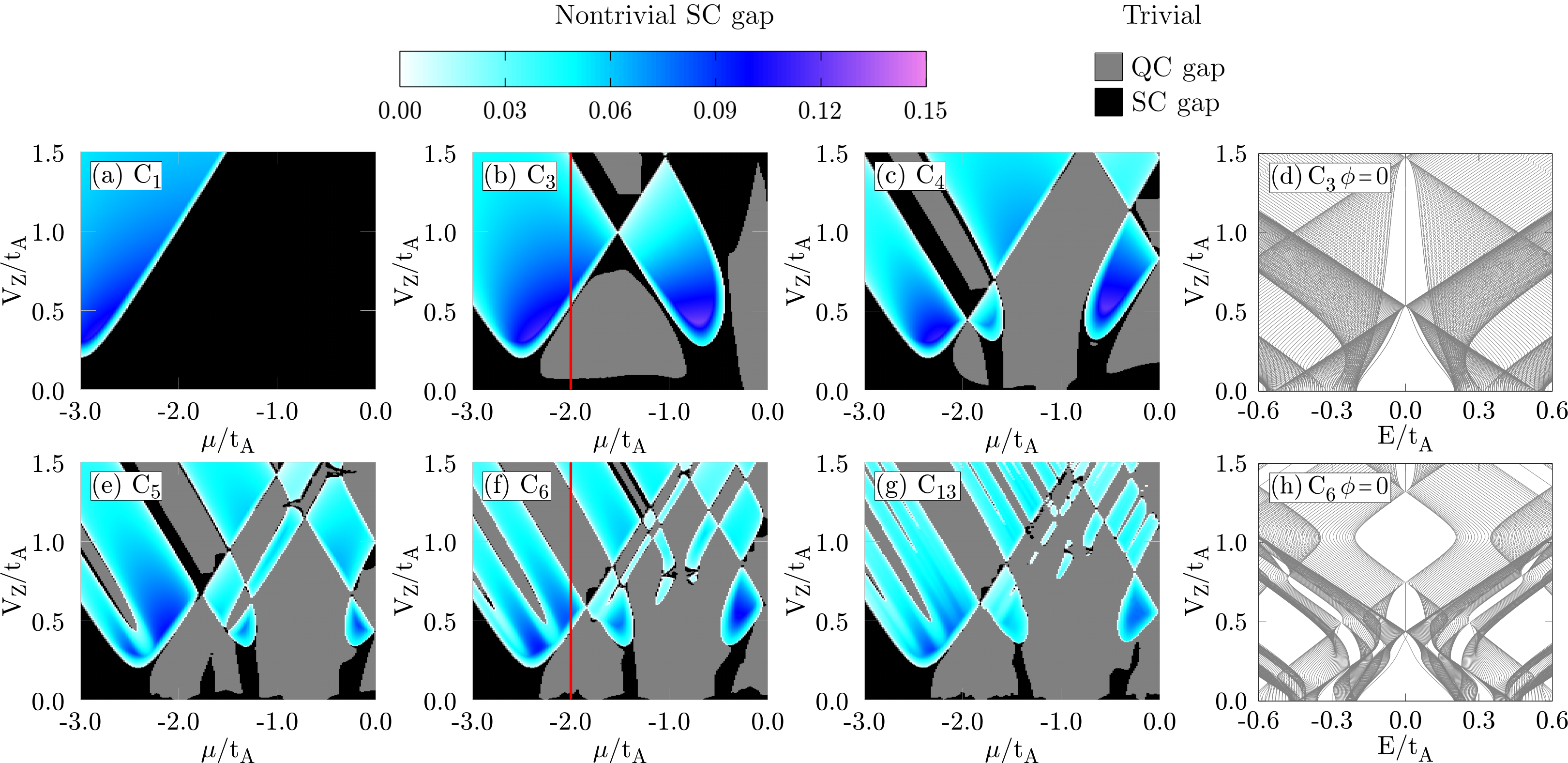}
\caption{
Phase diagrams as a function of chemical potential $\mu$ and magnetic field $\BZ$ for $C_1$, $C_3$, and $C_4$ (top row), and  $C_5$, $C_6$, and $C_{13}$ (bottom row) approximants.
Panels (d,h) show the energy spectrum as a function of $\BZ$ for $C_3$ and $C_6$ at the red lines in (b,f), respectively.
Parameters used: $\rho=1.5$, $\Delta=0.2t_A$, $\alpha=0.2t_A$. Panels (a,d,h) were generated for $\phi=0$, while the rest of the panels are generated for all relevant phason angles $\phi$ that force the QC subgap state to wind around the gap.
}
\label{fig:map_mB}
\end{figure*}
%
%%%%%%%%%%%%%%%%%%%%%%%%%%%%%%%%%%%%%%%%%%%%%%%%%%%%%%
%%%%%%%%%%%%%%%%%% B-DEPENDENCE%%%%%%%%%%%%%%%%%%%%%%%
%%%%%%%%%%%%%%%%%%%%%%%%%%%%%%%%%%%%%%%%%%%%%%%%%%%%%%
\subsection{Magnetic field dependence}
\label{sec.res.magnetic_field}
As shown in the previous subsection, the hopping ratio $\rho$ is an effective variable to tune the importance of quasiperiodicity. 
However, from an experimental standpoint it could be difficult to vary this parameter as it likely depends on both the type of substrate and the angle of the atomic chain with respect to the crystal lattice unit axis.
Therefore, in this subsection, we further quantify the influence of the magnetic field on the topological phase diagram, which can be varied by either the type of atoms used to produce the FC, tuning the effective exchange interaction, or by an applied external magnetic field~\cite{gierz.meier.11,bialo.23}.
Specifically, we compute the topological phase diagram as a function of the magnetic field $\BZ$ and the chemical potential $\mu$ in Fig.~\ref{fig:map_mB} using the same layout as Fig.~\ref{fig:map_mr}, now fixing $\rho = 1.5$.
Here, the topological phase diagram for $C_1$ FC (a) shows the typical scenario of the topological phase emerging, as described by Eq.~(\ref{eq:critical_B}) (due to the change in bandwidth scaling with $t_B$, the topological phase is shifted to lower values of $\mu$).

For the first real QC approximant, the $C_3$ approximant in Fig.~\ref{fig:map_mB}(b), a large satellite topological phase emerges in a parameter regime that is topologically trivial for a crystalline atomic chain. 
It is the competition between quasiperiodicity and topological superconductivity that influences the condition for the value of the critical magnetic field and thereby allows for emergence of satellite nontrivial MBS phases.
This is a direct result of the QC gap closings, leading to new gap openings that can be topologically nontrivial and host MBS.
An analytical description of this condition is usually obtained by finding the gap closings~\cite{sato.takahashi.10}, but here we can just modify the crystalline condition given in Eq.~(\ref{eq:critical_B}) by taking into account the FC structure of $C_3$. This results in four solutions for the gap closings and openings
\begin{numcases}
{ V_\pm^{C_3}=}
 \label{eq:critical_B:C_3} \nonumber 
 \left[\Delta^2+ \left(t_{\rm A} + t_{\rm B} \pm \mu \right)^2\right]^{\frac{1}{2}} ,& \\  
 \left[\Delta ^2+\mu ^2+\left(t_{\rm A} -t_{\rm B}\right)^2 - 4 \alpha^2 \right. \\
 \left.\pm 2 \sqrt{\left(\mu ^2 -4 \alpha ^2 \right) \left( t_{\rm A} - t_{\rm B}\right)^2 -4 \alpha ^2\Delta ^2}\right]^{\frac{1}{2}}. \nonumber
\end{numcases} 
This extension to $C_3$, and thus the inclusion of a secondary type of hopping, allows for additional solutions of this gap closing equation, which is the formal reason behind the satellite nontrivial topological phases.
Importantly, the satellite phase at high chemical potential hosts the largest topological gap, thereby generating the best robustness of the nontrivial phase. 
Moreover, for the studied range of parameters, the FC composed of $C_3$ approximants covers the highest percentage of the  phase diagram with the nontrivial phase hosting MBS among all approximants we investigate. 
We further observe the same pattern here as in Fig.~\ref{fig:map_mr}, that a trivial SC phase acts as a buffer between the QC and MBS phases and that this buffer also acts as an effective precursor for the MBS phase, as it removes the QC subgap states.

Moving on to the $C_4$ approximant in Fig.~\ref{fig:map_mB}(c), we find additional satellite nontrivial MBS regions. In a similar manner to Eq.~(\ref{eq:critical_B:C_3}), we can extend the condition for the gap closings to FC composed of $C_4$ approximants, now given by the six solutions of the cubic equation
\begin{equation}
    (V_\pm^{C_4})^3 + a (V_\pm^{C_4})^2 + b_\pm V_\pm^{C_4} + c_\pm = 0,
    \label{eq:critical_B:C_4}
\end{equation}
where the coefficients $a,\,b_{\pm},\, {\rm and },\,c_{\pm}$ depend on $\alpha,\,\mu,\,\rho$ and $\Delta$, see Appendix~\ref{app.c4.polynomial} for the full results. It is this proliferation of solutions to the gap equation that generates the increased number of topological regions. These analytically obtained topological boundaries for the $C_4$ approximant compare well with the numerical results. 
For even higher order approximants, Fig.~\ref{fig:map_mB}(e-g), the overall outline of the topological regions is rather similar, except that each region becomes ever more fragmented due to closely overlapping nontrivial regions and a multitude of QC gaps. 
Overall, this fragmentation suppresses the topological gap, such that higher order approximants have generally smaller gaps and the largest gaps are then often also found in the original, crystalline, topological region and not in the satellite regions.

Finally, in Figs.~\ref{fig:map_mB}(d,h) we show the energy spectrum extracted for the phason angle $\phi=0$, along the red lines of in parameter space in (b,f), respectively. At $\BZ=0$, there are no subgap states and the system is in the trivial SC phase. 
With increasing magnetic field, two subgap states appear, signifying a QC phase up until the transition to the nontrivial MBS phase around $\BZ \sim 0.5t_A$. 
Then, for the $C_3$ approximant (d), this nontrivial phase continues until very high magnetic fields. 
In contrast, for the $C_6$ approximant (h), an additional QC phase appears at zero energy as a QC gap is incident on the nontrivial gap and wins in the ensuing competition due to its larger size [c.f.~Fig.~\ref{fig:Spectra}(c)]. At $\phi=0$ its subgap states however resides at the gap edge.

For the sake of completeness, we also extract the phase diagram as a function of hopping ratio $\rho$ and magnetic field $\BZ$, see Appendix~\ref{app.V_rho}.
We find that quasiperiodicity again has a beneficial impact in the sense of a nontrivial phase appearing over a wider range of parameters compared to the crystalline limit, in particular for shorter approximants.

Summarizing the results for the topological phase diagram, we clearly observe a beneficial influence of the Fibonacci chains on the nontrivial MBS phase.
The topological MBS phase appears in a wider range of parameters, especially as satellite regions in parameter regimes where the crystalline chain can never be topologically nontrivial.
The size of the topological gap $\DeltaT$ can also be larger than that of the crystalline atomic chain.
However, for higher order approximants an increased fragmentation occurs in the topological regions. 
Thus, even though the topological regions can be rather large, the topological gap becomes suppressed for higher order approximants.
Therefore, as a conjecture, an infinitely long QC magnetic chain will produce a phase diagram with a myriad of infinitesimally small topological phases. 
There, even a slight change in parameters would shift the system from the topologically nontrivial to trivial phase, making it effectively impossible to verify the existence of MBS.% and might also lead to unforeseen consequences.
All of these results point to the benefit of short approximants, such as $C_{3,4}$, for generating topological superconductivity using chains of magnetic atoms on a superconducting substrate.

\section{Quasiperiodic SOC}
\label{sec.res.soc}
So far we have assumed a constant SOC $\alpha$, as described in Sec.~\ref{sec.model.tsc}. It might also be realistic to consider a SOC that changes proportionally to the hopping in the QC.
In this section, we therefore compare the already extracted topological phase diagrams with a model where the SOC changes proportionally to the bond length, here scaled with $t_{A,B}$, i.e.~we use $\alpha_B = \alpha t_B / t_A = \alpha \rho$ and $\alpha_A = \alpha$, on B and A bonds, respectively.
In other words, this corresponds to a quasiperiodic SOC with the same spatial modulation as the Fibonacci hopping terms.

\begin{figure*}[t]
\includegraphics[width=\linewidth]{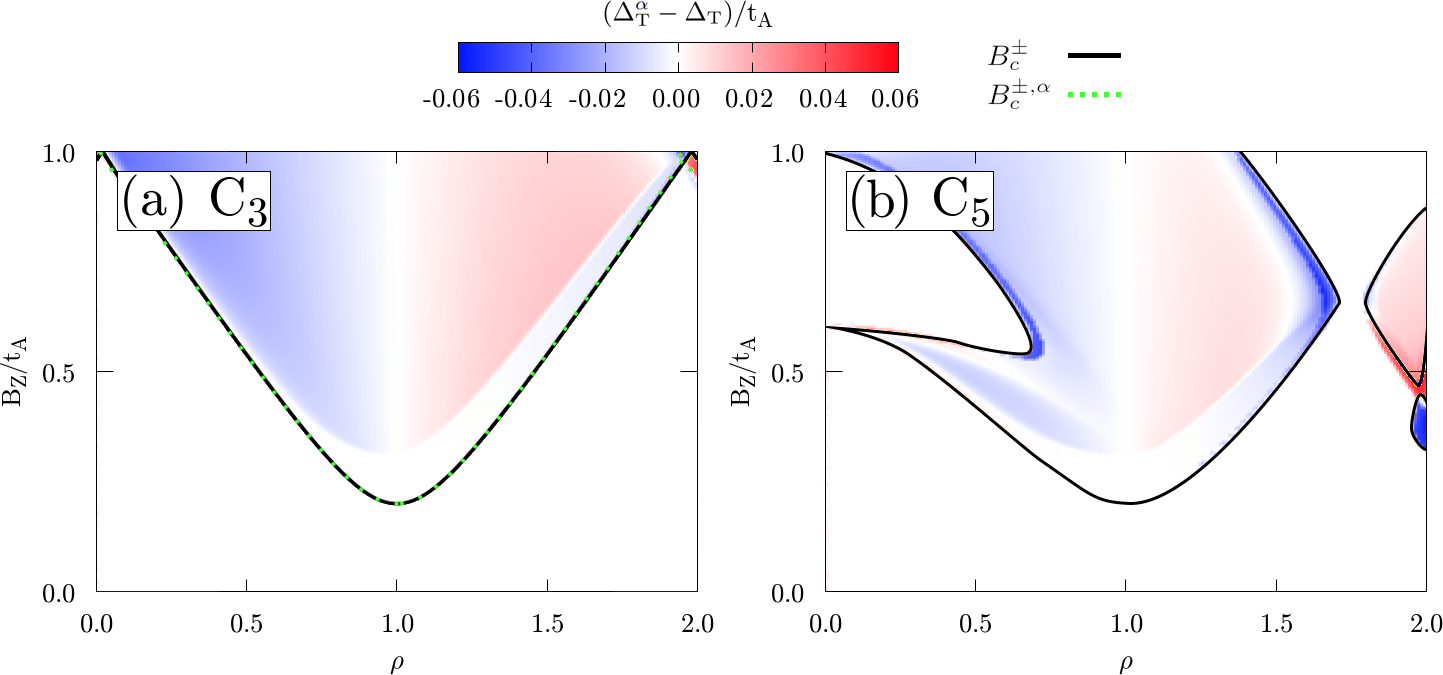}
\caption{Difference between the MBS gap sizes $\DeltaTalpha$ and $\DeltaT$ obtained with quasiperiodic SOC and constant SOC, respectively, as a function of hopping ratio $\rho$ and magnetic field $\BZ$ for approximants $C_3$ (a) and $C_5$ (b).
Black solid (green dashed) line on (a) denotes $\Bcpm$ ($\BcpmAlpha$) for constant (quasiperiodic) SOC.
Parameters used: $\Delta=0.2t_A$, $\alpha=0.2t_A$, $\mu=-2t_A$. 
}
\label{fig:map_alpha}
\end{figure*}

We start by studying the influence of the quasiperiodic SOC on the topological phase transition, both analytically and numerically. 
By generalizing the expression for the critical field $\Bcpm$ in the same manner as previously, we find that the quasiperiodic SOC modifies Eq.~(\ref{eq:critical_B:C_3}) for $C_3$ to become
\begin{equation}
\label{eq:critical_B:soc}
\begin{split}
&V_{\pm,\alpha}^{C_3} =  \left[\Delta^2 + \mu^2 + \left(t_{\rm A} \pm t_{\rm B}\right)^2 - \alpha^2\left(t_{\rm A} \mp t_{\rm B} \right)^2+ \right. \\
& \left.   2 \eta \sqrt{\mu ^2 \left(t_{\rm A} \pm t_{\rm B} \right)^2-\alpha ^2 \left[\Delta ^2 + \left(t_{\rm A}\pm t_{\rm B} \right)^2 \right] \left(t_{\rm A} \mp t_{\rm B} \right)^2}
 \right]^{\frac{1}{2}},
\end{split}
\end{equation}
where $\eta=\pm 1$ yields four solutions in total.
However, we find that this equation produces a nearly identical critical field as the constant SOC situation (less than a percent difference for typical parameter values), which can be explained by the new terms scaling with $(\alpha/t_A)^2 \approx 10^{-4}$, which is typically very small compared to other relevant terms. 
Thus, when computing the topological phase diagram numerically, we find that the two models for SOC produce nearly identical topological regions as already found in e.g.~Figs.~\ref{fig:map_mr},~\ref{fig:map_mB}, and~\ref{fig:map_rB}.

Still, we find a notable influence of quasiperiodic SOC on the topological MBS gap size, which we denote $\DeltaTalpha$ for the quasiperiodic SOC.
To clearly illustrate this increase, we plot the gap difference $\DeltaTalpha - \DeltaT$ in Fig.~\ref{fig:map_alpha} as a function of hopping ratio $\rho$ and magnetic field $\BZ$, where (a) and (b) show Fibonacci approximants $C_3$ and $C_5$, respectively. 
The black solid line outlines the topological phase boundary obtained with constant SOC, extracted from Figs.~\ref{fig:map_rB}(b,e), and we find that this overlaps nearly perfectly with the region obtained using quasiperiodic SOC (green dashed line).
In Fig.~\ref{fig:map_alpha} blue (red) color denotes where a quasiperiodic (constant) SOC yields the larger topological gap and therefore increases topological robustness.
We here omit to quantify any gap difference in the trivial region, since we do not find any significant difference. 

Interestingly, Fig.~\ref{fig:map_alpha} uncovers that the topological MBS region is composed of two parts: one near the topological phase transition threshold, which is left unchanged (white region just above the black line) and one where the gap size is susceptible to varying SOC (red and blue colors).
The unchanging part corresponds to where the MBS gap is just opening, which is mainly governed by the magnetic field $\BZ$, and hence the SOC has no effect here. Instead, the SOC governs the energy gap deeper into the topological region, as also illustrated by the linear versus non-linear parts of the gap above $\BZ > \Bcp$ in Fig.~\ref{fig:Spectra}(a).
For the region where the size of the topological gap $\DeltaT$ is influenced by quasiperiodic SOC, we find that as the quasiperiodicity gets stronger in either direction (i.e.~further away from the crystalline $\rho=1$) there is an increased difference in the topological gap size. Specifically, constant SOC (quasiperodic SOC) yields a stronger topological robustness for $\rho<1$ ($\rho>1$). 
Thus, we observe that the introduction of a quasiperiodic SOC strengthens the influence of quasiperiodicity on the topological gap. 
Additionally, its dependence on $\rho$ further increases the QC gap enhancement monotonically (up until $\rho \approx 1.6$).
The satellite topological regions follow a similar behavior, with an overall increased size of the topological gap and thus stronger MBS robustness for this set of parameters.
Finally, we comment briefly on results in higher order Fibonacci approximants. 
While the overall shapes of the topological regions are very similar between the two SOC models, the introduction of a quasiperiodic SOC allows for the existence of topological phases at slightly different parameters near the extreme values of $\rho \to 0$ or $2$. 
As the order of approximant increases, the fragmented nature of the topological regions also increases.
In conclusion, our main finding is that the gap enhancing effects of quasiperiodicity can increase even more with a quasiperiodic SOC. This result further establishes quasiperiodicity as useful for MBS formation.

%%%%%%%%%%%%%%%%%%%%%%%%%%%%%%%%%%%%%%%%%%%%%%%%%%
\section{Conclusions}
\label{sec.sum}
The experimental elusiveness of MBS motivates an extended search for an efficient and capable base structure at which their existence are favored. 
Thanks to recent progress in the fabrication of atomic chains on superconducting surfaces \cite{kim.palaciomorales.18,choi.lorente.19}, we here propose to utilize engineered quasiperiodic structures of magnetic atoms on a superconducting substrate to improve the possibilities for topological superconductivity with MBS. 
By simulating a quasiperiodic structure in the form of simple Fibonacci approximants, we find that quasiperiodicity leads to the emergence of additional large topological phase regions in parts of parameter space where any crystalline chain is topologically trivial. 
We also find that the topological gap, protecting the MBS, can be increased thanks to quasiperiodicity. The gap can be further increased by a quasiperiodic SOC. 
The optimal quasiperiodicity is reached for rather short QC approximants, such as $C_3$ and $C_4$, which is advantageous for experimental implementation. 
In contrast, higher order approximants generate too fragmented topological phases, with suppressed energy gaps.
At the same time, we find that MBS can only emerge in regions which are not depleted by QC gaps and that the topological superconducting state with MBS is mutually exclusive to the topological QC phase. 
In fact, while each of the QC gaps possesses a subgap state that winds across the QC gap with change of the phason angle $\phi$, the MBS in contrast show no signs of such winding or any other quasiperiodic properties, such as wave-function criticality. 

In conclusion, our work promotes quasiperiodic systems as a promising platform for studying both the rich interplay between the different topological phases and facilitating the emergence of MBS by triggering a number of additional topological phase transitions and even enhancing the robustness of MBS in some of these phases.

%%%%%%%%%%%%%%%%%%%%%%%%%%%%%%%%%%%%%%%%%%%%%%%%%%

%%%%%%%%%%%%%%%%%%%%%%%%%%%%%%%%%%%%%%%%%%%%%%%%%%
\begin{acknowledgments}
We thank P.~Dutta, K.~Tanaka, and A.~Sandberg for valuable discussions.
A.K., P.H., and A.M.B.S. acknowledge financial support from the Swedish Research Council (Vetenskapsr{\aa}det) Grant No.~2022-03963, and the Knut and Alice Wallenberg Foundation through the Wallenberg Academy Fellows program, KAW 2019.0309.
T.D. acknowledges the support of the National Science Center in Poland under grant No.  2022/45/B/ST3/02826.
O.A.A. and M.L. acknowledge funding from NanoLund, the Swedish Research Council (Grant Agreement No.~2020-03412) and the European Research Council (ERC) under the European Union's Horizon 2020 research and innovation programme under the Grant Agreement No.~856526.
The computations were enabled by resources provided by the National Academic Infrastructure for Supercomputing in Sweden (NAISS) and the Swedish National Infrastructure for Computing (SNIC) at UPPMAX, PDC and NSC, partially funded by the Swedish Research Council through grant agreements No.~2022-06725 and No.~2018-05973.
\end{acknowledgments}

%%%%%%%%%%%%%%%%%%%%%%%%%%%%%%%%%%%%%%%%%%%%%%%%%%
\appendix
\renewcommand{\thefigure}{A\arabic{figure}}
\setcounter{figure}{0}

\section{Topological classification using the Wilson loop}
\label{app.res.wilson}
In order to provide additional support for the classification of the nontrivial topological superconducting state hosting MBS in the main text, we show in this Appendix that we arrive at the same conclusion regarding the topological characterization when employing a completely different method based on the Wilson loop~\cite{Wang2019Band,Awoga2022Robust}.
\begin{figure}[b]
	\centering
\includegraphics[width=\linewidth]{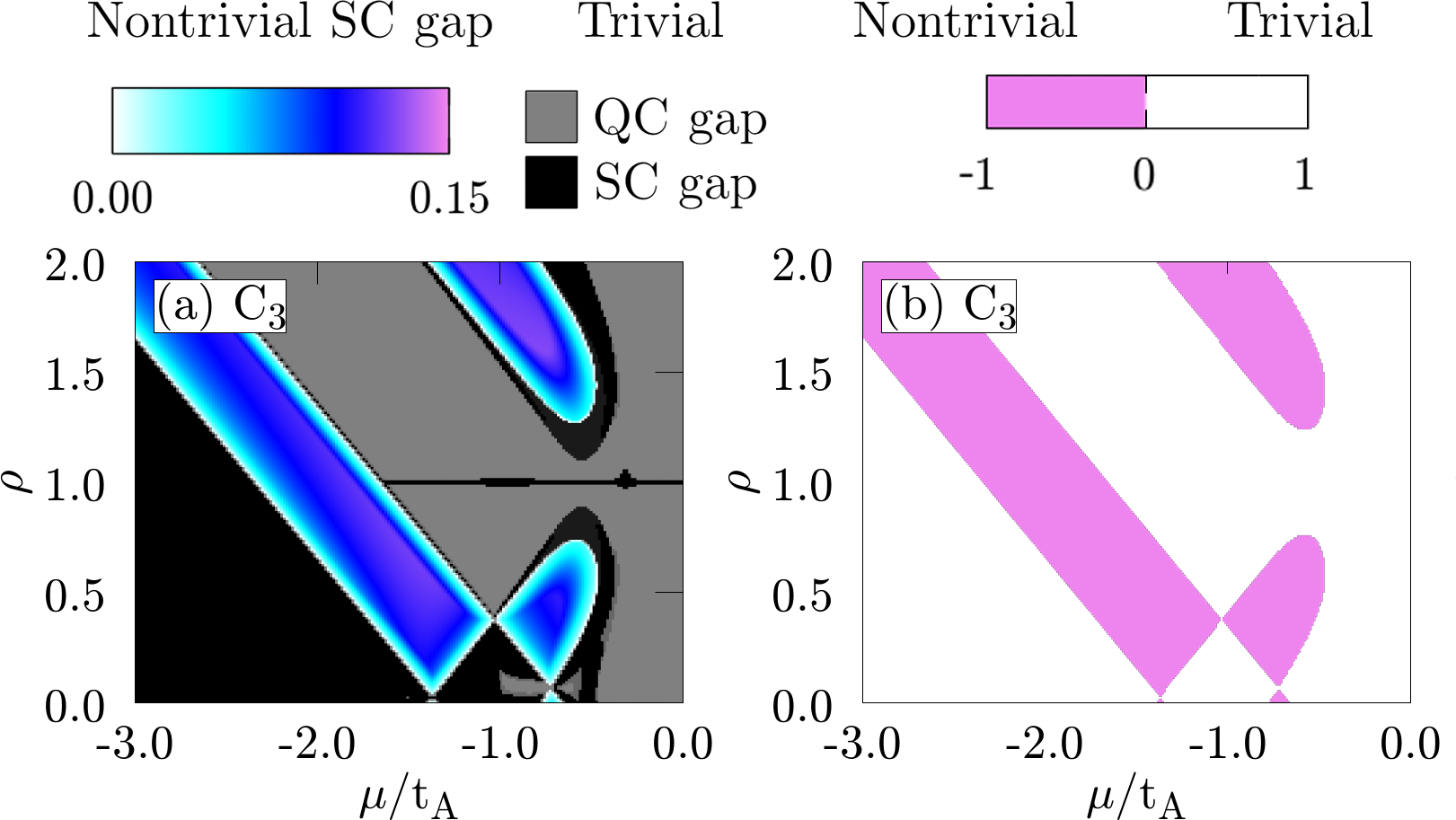}
	\caption{Comparison between methods for obtaining the topological phase diagram for the $C_3$ approximant: (a) method based on MP used in the main text and (b) Wilson loop method.
 } 
	\label{fig:comp}
\end{figure}
Specifically, in our analysis in the main text we use the MP to identify the topological MBS phase.
The Wilson loop is instead a geometric quantity with its argument being the Zak phase in 1D systems~\cite{Zak1989Berry}. 

To use the Wilson loop approach we consider an infinite repetition of the $n$:th Fibonacci approximant such that the chain forms a 1D superlattice with unit cell $F_n$. 
Then the Wilson loop $W$ is obtained as
\begin{equation}\label{eq:TopInv}
\begin{split}
 W & = \det \left[ \prod_{\substack{j=0\\k_j \in \rm BZ}}^{m-1} \hat{U}_{\rm o}\left( k_{j+1}\right)^\dagger \hat{U}_{\rm o}\left( k_j\right)\right] = e^{i\lambda},  
 \end{split}
\end{equation}
where $W=+1\left(-1\right)$ implies a topologically trivial (nontrivial) phase. 
Here, $\hat{U}_{\rm o}$ is a matrix of the eigenvectors (wave functions) of the occupied states, and it is a function of $k$, with $k_j \in \left[-\pi,\pi\right]$ the discretized value of $k$ into $m$ points. In addition, $\lambda$ is the Zak phase.
Since the wave function is defined up to a $U(1)$ phase, to ensure a gauge invariant $W$ we let $k_m \rightarrow k_0$ leading to $\hat{U}_{\rm o}\left(\pi\right) \rightarrow \hat{U}_{\rm o}\left(-\pi\right)$.
This procedure at the edge of the Brillouin zone eliminates any extra phase without influencing the value of $W$.
In Fig.~\ref{fig:comp} we present a comparative analysis of the phase diagrams of the $C_3$ approximant in $\mu - \rho$ parameter space, based on the method in the main text (a) and using the Wilson loop (b).
It is evident that the boundaries between the trivial and nontrivial phases are consistent across all phases, confirming the MP as a reliable indicator of MBS existence.
We also verify this across higher order approximants and for an extended parameter space, finding a full agreement.
In summary, we find that the topological classification based on the MP in the main text is consistent with the Wilson loop approach and thus accurately indicates the topological MBS phase.
\begin{figure*}
\includegraphics[width=\linewidth]{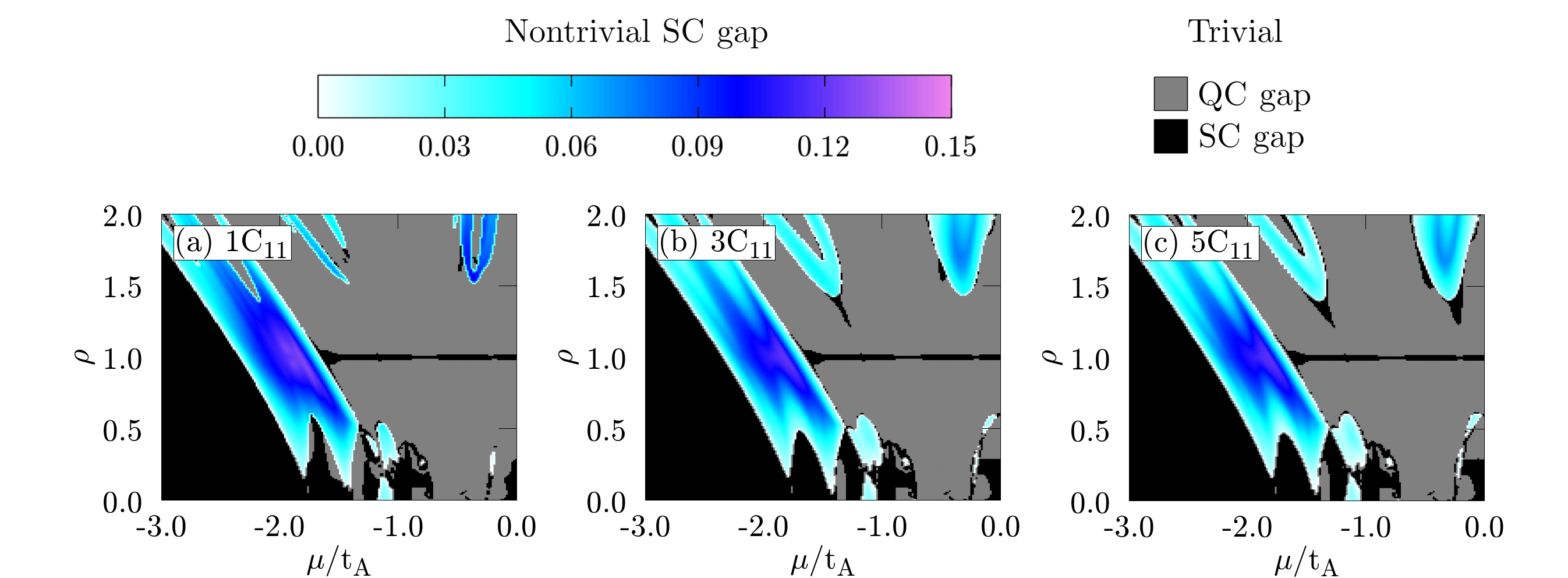}
\caption{
Topological phase diagram as a function of chemical potential $\mu$ and hopping ratio $\rho$ for the approximant $C_{11}$ repeated (a) $R=1$, (b) $R=3$, and (c) $R=5$ times. Colors indicate size of MBS gap, while grey (black) are gapped regions without MBS and with (without) other QC subgap states.
Parameters used: $\Delta=0.2t_A$, $\alpha=0.2t_A$, $\BZ=0.4t_A$.}
\label{fig:repeat}
\end{figure*}
\section{Influence of repetition of an approximant on the topological phase}
\label{app.repetition}

\begin{figure}[b]
\includegraphics[width=\linewidth]{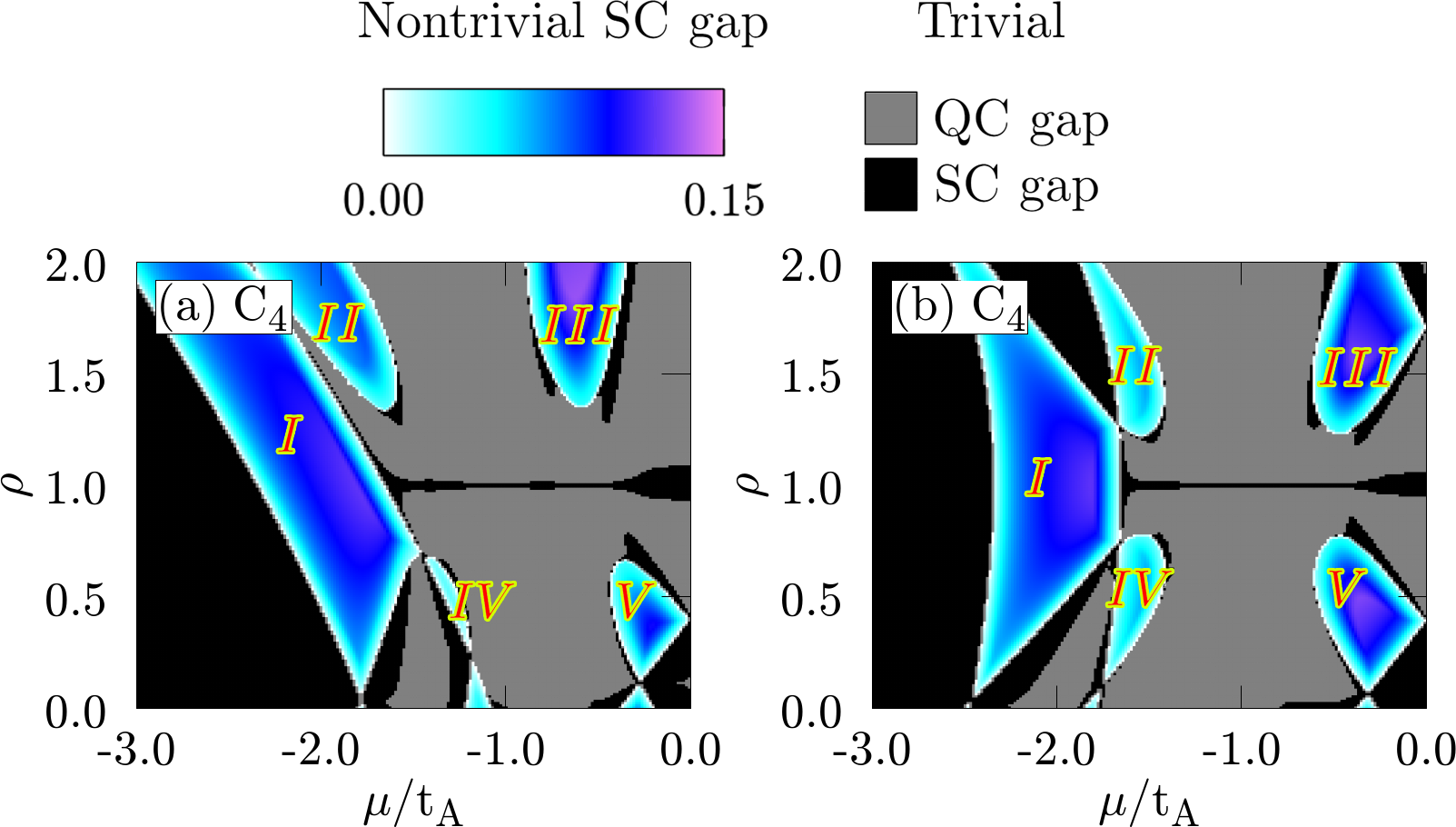}
\caption{Topological phase diagram as a function of chemical potential $\mu$ and hopping ratio $\rho$ for the approximant $C_4$ with bandwidth (a) not conserved and (b) conserved. 
Roman numerals $I-V$ depict the same nontrivial phases in the two cases.
Parameters used: $\BZ=0.4t_A$, $\Delta=0.2t_A$, 
 $\alpha=0.2t_A$.}
\label{fig:map_mrBw}
\end{figure}

\begin{figure*}[!t]
\includegraphics[width=\linewidth]{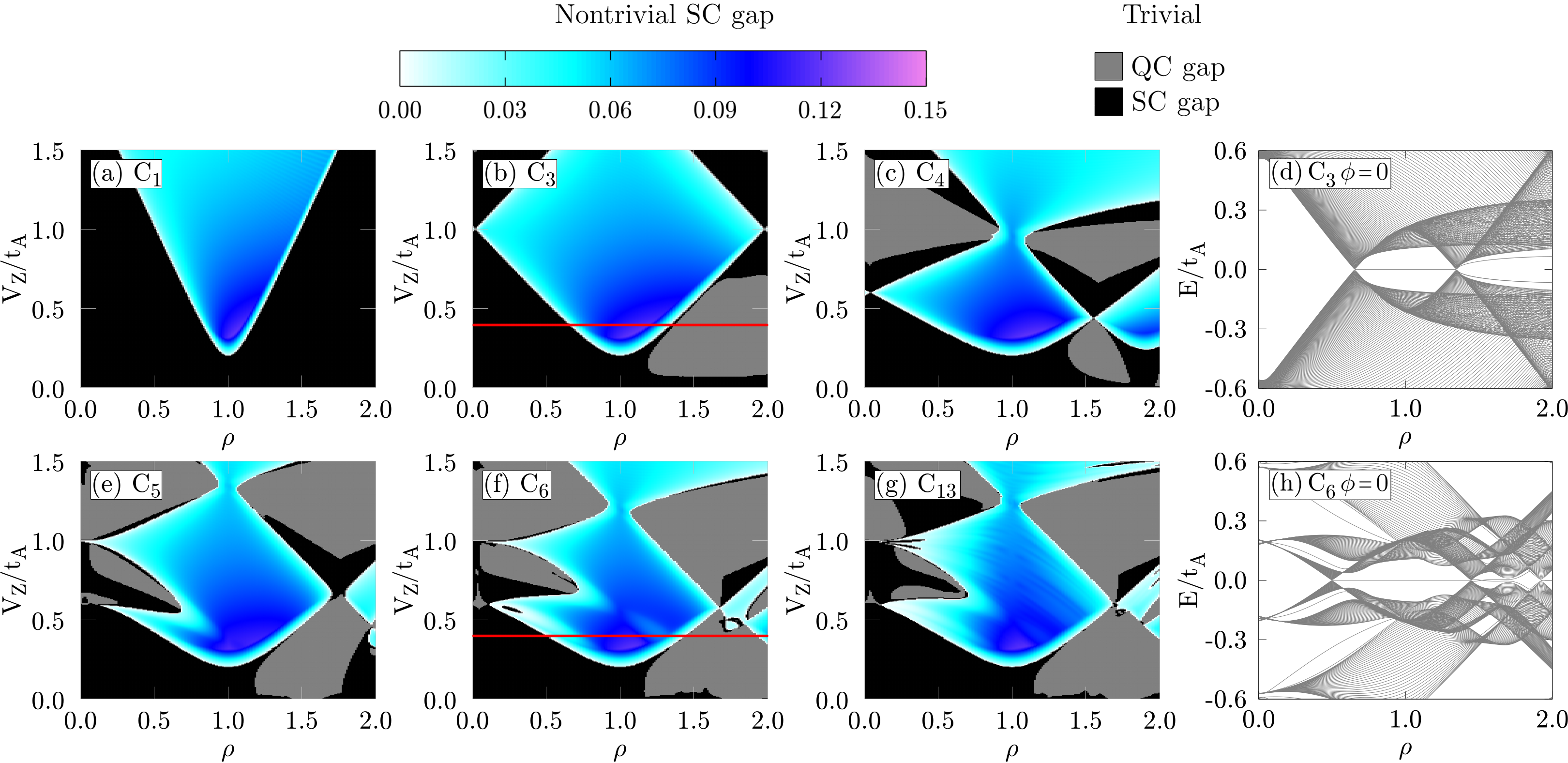}
\caption{
Phase diagrams as a function of hopping ratio $\rho$ and magnetic field $\BZ$ for $C_1$, $C_3$, and $C_4$ (top row), and  $C_5$, $C_6$, and $C_{13}$ (bottom row) approximants.
Panels (d,h) show the energy spectrum as a function of $\BZ$ for $C_3$ and $C_6$ indicated by red lines in (b,f) respectively. 
Parameters used: $\mu=-2t_A$, $\Delta=0.2t_A$, $\alpha=0.2t_A$. Panels (a,d,h) were generated for $\phi=0$, while the rest of the panels are generated for all relevant phason angles $\phi$ that force the QC subgap state to wind around the gap.}
\label{fig:map_rB}
\end{figure*}

In this Appendix we quantify how the repetition $R$ of the Fibonacci approximant influences the topological phase diagram studied in the main text, showing that the most important results are not influenced, as long as the MBS overlap is sufficiently small by having chosen $\Omega$ to be long enough.

We start by commenting on the system-size dependence in an atomic chain without quasiperiodicity, since we can understand all emergent results from this dependence.
It is well-known that MBS formation becomes robust once the wave function overlap between the MBS appearing at each chain end becomes negligible~\cite{cayao.BS.18}.
In the case of a finite Fibonacci approximant, repeating the approximant does not influence the major QC gap structure of the FC, but only reduces the level spacing in the non-gapped part of the spectrum~\cite{jagannathan.21,rai.21}. 
Thus, we do not expect the repetition to influence the topological phase diagram either, as long as the total system length is long enough to ensure vanishingly small MBS overlap.

In Fig.~\ref{fig:repeat} we show the $C_{11}$ approximant repeated (a) $R=1$, (b) $R=3$, and (c) $R=5$ times. 
The topological phases are shown as a function of the chemical potential $\mu$ and hopping ratio $\rho$, e.g.~to be compared with Fig.~\ref{fig:map_mr}(g) in the main text where the results for a very similar approximant $C_{13}$ ($R = 1$) are shown.
We see in Fig.~\ref{fig:repeat} that the overall shape of the topological regions is already established for $R=1$ and does not change substantially with $R$. 
Thus, we can be sure that the black buffer region encompassing the nontrivial phase will not fill out the nontrivial phase as the length of the atomic chain increases. 
Some minor changes are however seen in the satellite topological MBS regions, where the topological phase becomes more prominent for longer chains.
Moreover, an increased $R$ also decreases the MBS oscillations stemming from wavefunction overlap (not shown). 
To conclude, we find that increasing $R$ is generally beneficial for the topological MBS phase.

\section{Quasiperiodicity with conserved bandwidth}
\label{app.bandwidth}

In the main text, we use a QC hopping model that is commonly employed in quasiperiodic systems~\cite{jagannathan.21}, where the quasiperiodicity is quantified by the hopping ratio $\rho = t_B/t_A$. One consequence of this choice is that the bandwidth changes with $\rho$, which might in principle change the topological phase diagram.
Still, we point out that our most important results are obtained for both $\rho<1$ and $\rho>1$, i.e.~both for a smaller and larger bandwidth, respectively. 
Thus, our results must reasonably be related to quasiperiodicity, rather than bandwidth size.
For the sake of full transparency, we show in this Appendix that the topological phase diagrams obtained in Sec.~\ref{sec.res.phase_diagram} are indeed qualitatively the same when using a model where the bandwidth is conserved. 

We can preserve the bandwidth by modifying the $t_A$ and $t_B$ hoppings according to~\cite{rai.21}
\begin{align}
\label{eq:bw}
    F_{n-1}t_A+F_{n-2}t_B=F_n.
\end{align}
We present in Fig.~\ref{fig:map_mrBw} a comparison of the topological phase diagram for the $C_4$ approximant as a function of the chemical potential $\mu$ and $\rho$ with the bandwidth (a) not conserved, i.e.~same as in the main text and (b) when it is conserved, using Eq.~\eqref{eq:bw}.
When conserving the bandwidth, we find that the topological phase diagram becomes nearly symmetric with respect to $\rho=1$ line.
However, the two cases are still qualitatively similar as each of the topological regions is represented on both panels, see Roman numerals, although each region is somewhat stretched and twisted due to the nonlinear mapping of $\rho$ between both cases.
These results confirm that the topological regions are qualitatively comparable and portray the same phenomena, regardless of the implementation of quasiperiodicity and its related bandwidth.

\section{Critical field coefficients for approximant $C_4$}
\label{app.c4.polynomial}
In the main text we show in Eq.~(\ref{eq:critical_B:C_4}) the expression for the critical magnetic field for the bulk $s$-wave gap to close for the approximant $C_4$. Here we provide the coefficients,
\begin{align*}
 a  &=  -\left(4t_{\rm A}^2-6 \alpha ^2+3 \Delta ^2+3 \mu ^2+2 t_{\rm B}^2\right), \\
b_{\pm} & =  4 t_{\rm B}^4+ 9 \alpha^4+3 \Delta^4+3 \mu^4+4(t_{\rm A}^2 \pm 3 \mu t_{\rm A}) t_{\rm B}^2 \\
& + \Delta^2\left(4t_{\rm A}^2+6 \mu^2+8 t_{\rm B}^2\right) \\
&-2 \alpha^2\left(t_{\rm A}^2 + 8t_{\rm A}t_{\rm B} + 6 t_{\rm B}^2 +6 \mu^2 \pm 2 \mu(t_{\rm A}+2 t_{\rm B})\right) + t_{\rm A}^4,\\
c_{\pm} &= -\left[\alpha ^4 \left(9 \Delta ^2+((4t_{\rm A}\mp 3 \mu )+2 t_{\rm B} )^2\right) \right. \\
    & \left.  + 2 \alpha ^2 \left(3 \Delta ^4+3 \Delta ^2 \left(\left(t_{\rm A}^2+t_{\rm B}^2\mp 2 \mu  (2t_{\rm A} +t_{\rm B})\right)+2t_{\rm A}^2\right) \right. \right.\\
    & \left. \left. -(\mu  (\mu \mp  t_{\rm B})-2t_{\rm A}^2) (3 \mu \mp 2 (t_{\rm B} +2 t_{\rm A})) (\mu \pm t_{\rm B} )\right)  \right. \\
     & \left. +\left[\Delta ^4+\Delta ^2 \left(\left(2 \mu ^2\mp 2 \mu  t_{\rm B} \right)+ t_{\rm A}^2+4t_{\rm B}^2\right) \right. \right. \\
     &\left.\left. +(2t_{\rm A}^2 -\mu  (\mu \mp t_{\rm B} ))^2\right] \left[\Delta ^2+(\mu \pm t_{\rm B} )^2\right]\right]. 
\end{align*}

As seen, this represents a significantly more complicated solution compared to the $C_3$ approximant given in Eq.~(\ref{eq:critical_B:C_3}).
We refrain from solving for higher-order approximants, since the complexity quickly reduces the usefulness.
We note that this complexity is explicitly visualized through the fragmented topological phase diagrams in the higher order approximants.
\section{$\rho$--$\BZ$ topological phase diagram}
\label{app.V_rho}
In order to complete the investigation of the phase diagram in $\mu-\rho-\BZ$ space, we provide in this Appendix the topological phase diagram as a function of hopping ratio $\rho$ and magnetic field $\BZ$ in Fig.~\ref{fig:map_rB}, thus supplementing the discussion in Sec.~\ref{sec.res.magnetic_field}.
Here, the topological transition to the nontrivial phase occurs for the lowest magnetic field at $\rho=1$, which is a common point for each of the panels in this figure as such cross section portrays the same system, regardless of the Fibonacci approximant. 
Overall, we find similar results to those in Sec.~\ref{sec.res.magnetic_field}: the regions with a topological MBS phase increase and this phase can be found for parameters not available in a crystalline chain. Smaller approximants are generally more beneficial as the topological regions gets more fragmented with increasing lengths of the approximants. 
The topological gap also increases, especially for $\rho > 1$.

%%%%%%%%%%%%%%%%%%%%%%%%%%%%%%%%%%%%%%%%%%%%%%%%%%
\bibliography{biblio}
\end{document}